\documentclass[12pt,titlepage]{article}
\pdfoutput=1
\usepackage[footnotesize]{caption}
\usepackage{float}
\usepackage[symbol]{footmisc}
\usepackage{algorithm}
\usepackage{algpseudocode}
\usepackage[superscript,sort]{cite}
\usepackage{array}
\usepackage[pdftex,colorlinks=true,linkcolor=black,citecolor=black,urlcolor=black,bookmarks=true,breaklinks=true]{hyperref}
\usepackage[pdftex]{graphicx}
\usepackage[usenames,pdftex]{color}
\definecolor{Red}{rgb}{1.0,0.0,0.0}
\usepackage{amsmath,amssymb}
\usepackage{amsthm,amscd,amsxtra,amsfonts,amsmath,amssymb,multirow}
\usepackage{wrapfig}
\usepackage[tiny,compact]{titlesec}
\usepackage{helvet}
\usepackage{xcolor,colortbl}
\usepackage{subfigure}
\usepackage[T1]{fontenc}
\usepackage{times}
\DeclareMathAlphabet{\MATHIT}{OT1}{ptm}{m}{it}%% similiar to mathptmx
\DeclareSymbolFont{Letters}{OML}{ztmcm}{m}{it}%% dito
\DeclareSymbolFontAlphabet{\mathNormal}{Letters}% dito

\usepackage{longtable}
\usepackage{setspace}
\usepackage{fancyhdr}

\titlespacing*{\section}{0pt}{*0}{*0}
\titlespacing*{\subsection}{0pt}{*0}{*0}
\titlespacing*{\subsubsection}{0pt}{*0}{*0}
\titlespacing{\paragraph}{0pt}{*0}{*1}

\definecolor{MyPurple}{rgb}{1,0,1}

\newcommand{\beq}[1]{\begin{equation} \label{#1}}
\newcommand{\eeq}{\end{equation}}
\newcommand{\barray}{\begin{array}{ll}}
\newcommand{\earray}{\end{array}}

\makeatletter
\if@compatibility
\renewenvironment{titlepage} % changed from "newenvironment"
{%
\if@twocolumn
\@restonecoltrue\onecolumn
\else
\@restonecolfalse\newpage
\fi
\thispagestyle{headings}% Changed from "empty"
\setcounter{page}\z@
}%
{\if@restonecol\twocolumn \else \newpage \fi
}
\else
{%
\if@twocolumn
\@restonecoltrue\onecolumn
\else
\@restonecolfalse\newpage
\fi
\thispagestyle{fancy}% changed from "empty"
\setcounter{page}\@ne
}%
{\if@restonecol\twocolumn \else \newpage \fi
\if@twoside\else
\setcounter{page}\@ne
\fi
}
\fi
\makeatother

\begin{document}
\pagenumbering{roman}

\clearpage \pagebreak \setcounter{page}{1}
\renewcommand{\thepage}{{\arabic{page}}}

\title{A topological approach for protein classification}

\author{Zixuan Cang$^1$, Lin Mu$^2$, Kedi Wu$^1$,  Kristopher Opron$^3$,
Kelin Xia$^1$ and
Guo-Wei Wei$^{4}$ \footnote{ On leave from the Department of Mathematics, Michigan State University}~\footnote{ Address correspondences  to Guo-Wei Wei. E-mail:wei@math.msu.edu}\\
%\address{
$^1$ Department of Mathematics \\
Michigan State University, East Lansing, MI 48824, USA\\
$^2$ Oak Ridge National Laboratory\\
One Bethel Valley Road,
P.O. Box 2008, MS 6211, 
Oak Ridge, TN 37831, USA\\
$^3$  Department of Biochemistry and Molecular Biology\\
Michigan State University, East Lansing, MI 48824, USA \\
$^4$ Mathematical Biosciences Institute\\
The Ohio State University,
%Jennings Hall, 3rd Floor,  1735 Neil Avenue,
Columbus, Ohio 43210, USA
}

%\rhead{Topological protein classification}

\date{\today}
\pagenumbering{gobble}

\maketitle

\begin{abstract}
Protein function and dynamics are closely related to its sequence and structure. However prediction of protein function and dynamics from its sequence and structure is still a fundamental challenge in molecular biology. Protein classification, which is typically done through  measuring the similarity between proteins based on protein sequence or physical information,  serves as a crucial step toward the understanding of  protein function and dynamics. Persistent homology is a new branch of algebraic topology that  has found its success in the topological data analysis in  a variety of disciplines, including molecular biology. The present work explores the potential of using persistent homology as an independent tool for protein classification.   To this end, we propose a molecular topological fingerprint based support vector machine (MTF-SVM) classifier. Specifically,  we construct machine learning feature vectors solely from protein topological fingerprints, which are topological invariants generated during the filtration process. To validate the present MTF-SVM approach, we consider four types of problems. First, we study  protein-drug binding by using  the  M2 channel protein of  influenza A virus. We achieve  96\% accuracy in discriminating drug bound and unbound M2 channels.  Additionally, we examine the use of MTF-SVM for the classification of hemoglobin molecules in their relaxed and taut forms and obtain about 80\% accuracy. The  identification of all alpha, all beta, and alpha-beta protein domains is carried out in our next study using 900 proteins.  We have found a 85\% success in this  identification. Finally, we apply the present technique to 55 classification tasks of protein superfamilies over  1357 samples. An average accuracy of 82\% is attained. The present study establishes computational topology as an independent and effective alternative for protein classification.

\end{abstract}


\centerline{\bf Running title:  Topological  protein classification}

\maketitle

\pagebreak
{\setcounter{tocdepth}{4} \tableofcontents}
\pagebreak
 \setcounter{page}{1}
 \renewcommand{\thepage}{{\arabic{page}}}
\pagebreak

\section{Introduction}\label{sec:Intro}

% protein structure-function relation
Proteins are essential building blocks of living organisms. They function as catalyst, structural elements, chemical signals, receptors, etc.  The molecular mechanism of protein functions are closely related to their structures. The study of structure-function relationship is the  holy grail of biophysics and has attracted enormous effort in the past few decades. The understanding of such a relationship enables us to predict protein functions from structure or    amino acid sequence or both, which remains major challenge in molecular biology.  Intensive experimental investigation has been carried out to explore the interactions among proteins or proteins with other biomolecules, e.g., DNAs and/or RNAs. In particular, the understanding of protein-drug interactions is of premier importance to human health.

%biophysical approcahes
A wide variety of theoretical and computational approaches has been proposed to understand the protein structure-function relationship.  One class of approaches is biophysical. From the point of view of biophysics, protein structure, function, dynamics and transport are, in general, dictated by protein interactions. Quantum mechanics (QM) is based on the fundamental principle, and offers the most accurate description of interactions among electrons, photons, atoms and even molecules. Although QM methods have unveiled many underlying mechanisms of reaction kinetics and  enzymatic activities, they typically are computationally too expensive to do for large biomolecules. Based on  classic physical laws, molecular mechanics (MM) \cite{McCammon:1977} can, in combination with fitted parameters, simulate the physical movement of atoms or molecules for relatively large biomolecular systems like proteins quite precisely. However, it can be computationally intractable for macromoelcular systems involving realistic biological time scales.  Many time-independent methods like normal mode analysis (NMA)  \cite{Go:1983,Tasumi:1982,Brooks:1983,Levitt:1985}, elastic network model (ENM) \cite{Tirion:1996,Flory:1976,Bahar:1997,Lee:2007},  graph theory \cite{Jacobs:2001} and flexibility-rigidity index (FRI) \cite{KLXia:2013d,Opron:2014,Opron:2015a}  are proposed to capture features of large biomolecules.
Variational multiscale methods \cite{Wei:2009,Wei:2012,Wei:2013,ZhanChen:2010a,ZhanChen:2010b,ZhanChen:2012, DuanChen:2012a,DuanChen:2012b} are another class of approaches that combine atomistic description with continuum approximations.
There are well developed servers for predicting protein functions based on three-dimensional (3D) structures \cite{Laskowski:2005} or models from the homology modeling (here homology is in biological sense) of amino acid sequence if 3D structure is not yet available \cite{Roy:2010}. %Theoretical and semi-empirical approaches have been carried out to study the mechanism of proteins.
%There are well developed servers predicting protein function based on 3D structure \cite{Laskowski:2005} or models from homology modeling of amino acid sequence if 3D structure is not yet available \cite{Roy:2010}.

% Bioinformative approaches
Another class of important approaches, bioinformatical methods, plays a unique role for the understanding of the structure-function relationship. These data-driven predictions  are based on similarity analysis. The essential idea is that proteins with similar sequences or structures may share similar functions. Also, based on sequential or structural similarity, proteins can be classified into many different groups. Once the sequence or structure of a novel protein is identified, its function can be predicted by assigning it to the group of proteins that share similarities to a good extent. However, the degree of similarity depends on the criteria used to measure similarity or difference.  Many measurements are used to describe similarity between two protein samples. Typical approaches use either sequence or physical information, or both.  Among them,  sequence alignment can describe how closely the two proteins are related. Protein blast \cite{Johnson:2008}, clustalW2 \cite{Li:2015}, and other software packages can preform global or local sequence alignments. Based on sequence alignments, various scoring methods can provide the description of protein similarity \cite{Henikoff:1992, Altschul:1993}. Additionally, sequence features such as sequence length and occurrence percentage of a specific amino acid can also be employed to compare proteins. Many sequence based features can be derived from the position-specific scoring matrix (PSSM) \cite{Stormo:1982}. Moreover, structural information provides an efficient description of protein similarity as well. Structure alignment methods include rigid, flexible and other methods. The combination of different structure alignment methods and different measurements such as root-mean-square deviation (RMSD) and Z-score gives rise to various ways to quantify the similarity among proteins. As per structure information, different physical properties such as  surface area, volume, free energy, flexible-rigidity index (FRI) \cite{KLXia:2013d,Opron:2014,Opron:2015a}, curvature \cite{XFeng:2012a,XFeng:2013b}, electrostatics \cite{Zhou:2008b} etc.  can be calculated. A continuum model, Poisson Boltzmann (PB) equation delivers quite accurate estimation for electrostatics of biomolecules. There are many efficient and accurate PB solvers including PBEQ \cite{Jo:2008}, MIBPB \cite{Zhou:2008b,DuanChen:2011a}, etc. Together with physical properties, one can also extract geometrical properties from structure information. These properties include coordinates of atoms, connections between atoms such as covalent bonds and hydrogen bonds, molecular surfaces \cite{Bates:2008,Bates:2009,QZheng:2012} and curvatures \cite{XFeng:2012a,XFeng:2013b, WeiWu:2013}. These various approaches reveal information of different scales from local atom arrangement to global architecture. Physical and geometrical properties described above add different perspective to analyze protein similarities.

%Machine learning
Due to the advance in bioscience and biotechnology, biomolecular structure date sets are growing at an unprecedented rate. For example, the  \href{http://www.rcsb.org/pdb/home/home.do}{Protein Data Bank (PDB)} has accumulated more than a hundred thousand biomolecular structures.    The prediction of the protein structure-function relationship from such  huge amount of data can be extremely challenging. Additionally, an eve-growing number of physical or sequence  features are evaluated for each data set or amino-acid residue, which adds to the complexity of the data-driven prediction. To automatically analyze excessively large  data sets in molecular biology, many machine learning methods have been developed \cite{Leslie:2004,JLCheng:2006,Meinicke:2015,Fernandez-Lozano:2014}. These methods are mainly utilized for the classification, regression, comparison  and clustering of biomolecular data. Clustering is an unsupervised learning method which divides a set of inputs into groups without knowing the groups beforehand. This method can unveil  hidden patterns in the data set. Classification is a supervised learning method, in which, a classifier is trained on a given training set and used to do prediction for new observations. It assigns observation to one of several pre-determined categories based on knowledge from training data set in which the label of observations is known. Popular methods for classification include support vector machine (SVM) \cite{Burges:1998}, artificial neural network (ANN) \cite{McCulloch:1943}, deep learning \cite{Hinton:2006}, etc. In classification, each observation in training the set has a feature vector that describes the observation from various perspectives and a label that indicates to which group the observation belongs. A model trained on the training set indicates to which group  a new observation belongs with feature vector and unknown label. To improve the speed of classification and reduce effect from irrelevant features, many feature selection procedures have been proposed \cite{Dash:1997} .  Machine learning approach are successfully  used for protein hot spot prediction \cite{Darnell:2008}.

%Limiation of sequence and phsyical based approaches.
The data-driven analysis of the protein structure-function relationship is compromised by the fact that same protein may have different conformations which possess different properties or delivers different functions. For instance,   hemoglobins have  taut form with low affinity to oxygen and relaxed form with high affinity to oxygen; and ion channels often have open and close states.  Different conformations of a given protein  may only have minor differences in their local geometric configurations. These conformations share the same sequence and may have very similar physical properties. However, their minor structural differences might lead to dramatically different functions. Therefore, apart from the conventional physical and sequence information, geometric and topological information can also play an important role in understanding the protein structure-function relationship. Indeed, geometric information has been extensively used in the protein exploration.
  In contrast, topological information has been hardly employed in studying the protein structure-function relationship.

% persistent homology
In general, geometric approaches are frequently  inundated with too much geometric detail  and  are often   prohibitively expensive for most realistic biomolecular systems, while traditional topological methods often incur in too much reduction of the original geometric and physical information. Persistent homology, a new branch of applied topology, is able to bridge traditional geometry and topology.  It creates a variety of topologies of a given object by varying a  filtration parameter, such as a radius or a level set function. In the past decade, persistent homology has been developed as a new multiscale representation of topological features.  The 0-th dimensional version was originally introduced for computer vision applications under the name ``size function" \cite{Fro90, Frosini:1999} and the idea was also studied by Robins \cite{Robins:1999}. The Persistent homology theory was formulated, together with an algorithm given, by Edelsbrunner et al. \cite{Edelsbrunner:2002}, and a more general theory was developed by Zomorodian and Carlsson \cite{Zomorodian:2005}. There has since been significant theoretical development \cite{BH11,CEH07,CEH09,CEHM09,CCG09,CGOS11,Carlsson:2009theory,CSM09,SMV11,zigzag}, as well as various computational algorithms \cite{OS13,DFW14,Mischaikow:2013,javaPlex,Perseus, Dipha}. Often, persistent homology can be visualized through barcodes \cite{CZOG05,Ghrist:2008}, in which various horizontal line segments or bars are the homology generators which survive over filtration scales. Persistence diagrams are another equivalent representation \cite{edelsbrunner:2010}.  { Computational homology and persistent homology  have been applied to a variety of domains, including image analysis \cite{Carlsson:2008,Pachauri:2011,Singh:2008,Bendich:2010,Frosini:2013}, chaotic dynamics verification \cite{Mischaikow:1999,kaczynski:mischaikow:mrozek:04}, sensor network \cite{Silva:2005}, complex network \cite{LeeH:2012,Horak:2009}, data analysis \cite{Carlsson:2009,Niyogi:2011,BeiWang:2011,Rieck:2012,XuLiu:2012}, shape recognition \cite{DiFabio:2011,AEHW06} and computational biology \cite{Kasson:2007,Gameiro:2014,Dabaghian:2012,Perea:2015a,Perea:2015b}.} Compared with traditional computational topology \cite{Krishnamoorthy:2007,YaoY:2009,ChangHW:2013}  and/or computational homology, persistent homology  inherently has an additional dimension, the filtration parameter, which can be utilized to embed some crucial geometric or quantitative information into the topological invariants. The importance of retaining  geometric information in topological analysis has been recognized  \cite{Biasotti:2008}, and  topology has been advocated as a new approach for tackling  big data sets \cite{BVP15, BHPP14,Fujishiro:2000,Carlsson:2009,Ghrist:2008}.

% symmary of our work
Recently, we have introduced persistent homology for mathematical modeling and prediction of  nano particles, proteins and other biomolecules \cite{KLXia:2014c, KLXia:2015a}. We  have proposed  molecular topological fingerprint (MTF)   to reveal topology-function relationships in protein folding and protein flexibility \cite{KLXia:2014c}. We have employed  persistent homology to   predict  the curvature energies of fullerene isomers  \cite{KLXia:2015a,BaoWang:2014}, and analyze the stability of protein folding \cite{KLXia:2014c}.    More recently, we  have introduced resolution based persistent topology \cite{KLXia:2015d,KLXia:2015e}. Most recently, we have developed new multidimensional persistence,  a topic that has attracted much attention in the past few years \cite{Carlsson:2009computing,Carlsson:2009theory}, to better bridge geometry and traditional topology and achieve better characterization of biomolecular data \cite{KLXia:2015c}.  We have also introduced the use of topological fingerprint  for resolving ill-posed inverse problems in  cryo-EM  structure determination \cite{KLXia:2015b}.

 % objective
The objective of the present work is to explore the utility of MTFs for protein classification and analysis. We construct feature vectors based on MTFs to describe unique topological properties of protein in different scales, states and/or conformations. These topological feature vectors are further used in conjugation with the SVM algorithm for the classification of proteins. We validate the proposed MTF-SVM strategy  by  distinguishing different protein conformations, proteins with different local secondary structures, and proteins from different superfamilies or families. The performance of proposed topological method is demonstrated  by a number of realistic applications, including protein binding analysis, ion channel study, etc.

% layout of the paper
The rest of the paper is organized as following. Section \ref{sec:methods} is devoted to  the mathematical foundations for persistent homology and machine learning methods.
We present a brief description of simplex and simplicial complex followed by basic concept of homology, filtration, and persistence in Section \ref{Persistenthomology}. Three different methods to get simplicial complex, Vietoris-Rips complex, alpha complex, and \v{C}ech complex are discussed. We use a sequence of graphs of channel proteins  to illustrate the growth of a Vietoris-Rips complex and corresponding  barcode representation of topological persistence.     In Section \ref{svm+roc}, fundamental concept of support vector machine is discussed. An introduction   of transformation of the original optimization problem is given. A measurement for the performance of classification model known as receiver operating characteristic is described. Section \ref{feature+preprocessing} is devoted to the description of features used in the classification and pre-processing of topological feature vectors. In section \ref{sec:Numerical}, four test cases are shown. Case 1 and Case 2 examine the performance of the topological fingerprint  based classification methods in distinguishing different conformations of same proteins. In Case 1, we  use the structure of the M2 channel of influenza A virus with and without an inhibitor. In Case 2, we employ the structure of hemoglobin in taut form and relaxed form. Case 3 validates the proposed topological methods in capturing the difference between local secondary structures. In this study, proteins are divided into three groups, all alpha protein, all beta protein, and alpha+beta protein. In Case 4, the ability of the present method for distinguishing different protein families is examined. This paper ends with some concluding remarks.

\section{Materials and Methods}\label{sec:methods}

This section presents a brief review of persistent homology theory and illustrates its use in proteins. A brief description of machine learning methods is also given. The topological feature selection and construction from biomolecular data are described in details.

\subsection{Persistent homology}\label{Persistenthomology}
Points, edges, triangles and their higher dimensional counterparts are defined as simplices. A simplicial space is a topological space constructed from finitely many simplices.

\noindent
\textbf{Simplex} A $k$-simplex denoted by $\sigma^k$ is a convex hull of $k+1$ vertices which is represented by a set of points
\begin{equation}\label{simplex}
\sigma^k=\{\lambda_0u_0+\lambda_1u_1+...+\lambda_ku_k|\sum\lambda_i=1, \lambda_i\geq 0, i=0,1,...,k\},
\end{equation}
where $\{u_0,u_1,...,u_k\}\subset\mathbb{R}^n$ is a set of affinely independent points. Geometrically, a $1$-$simplex$ is a line segment, a $2$-simplex is a triangle, a $3$-simplex is a tetrahedron, and a $4$-simplex is a $5$-cell (a four dimensional object bounded by five tetrahedrons). A $m-$face of the $k$-simplex is defined as a convex hull formed from a subset consisting $m$ vertices.

\noindent
\textbf{Simplicial complex} A simplicial complex $\mathcal{K}$ is a finite collection of simplices satisfying two conditions. First, faces of a simplex in $\mathcal{K}$ are also in $\mathcal{K}$; Second, intersection of any two simplices in $\mathcal{K}$ is a face of both the simplices. The highest dimension of simplices in $\mathcal{K}$ determines dimension of $\mathcal{K}$.

\noindent
\textbf{Homology} For a simplicial complex $\mathcal{K}$, a $k$-chain is a formal sum of the form $\sum_{i=1}^Nc_i[\sigma_i^k]$, where $[\sigma_i^k]$ is oriented $k$-simplex from $\mathcal{K}$. For simplicity, we choose $c_i\in\mathbb{Z}_2$.  All these $k$-chains on $\mathcal{K}$ form an Abelian group, called chain group and denoted as $C_k(\mathcal{K})$. %Every dimension of a simplicial complex has a corresponding chain group.
A boundary operator $\partial_k$ over a $k$-simplex $\sigma^k$ is defined as,
\begin{equation}
\partial_k\sigma^k=\sum_{i=0}^{k}(-1)^i[ u_0,u_1,...,\widehat{u_i},...,u_k],
\end{equation}
where $[ u_0,u_1,...,\widehat{u_i},...,u_k ]$ denotes the face obtained by deleting the $i$th vertex in the simplex. The boundary operator induces a boundary homomorphism $\partial_k: C_k(\mathcal{K})\rightarrow C_{k-1}(\mathcal{K})$. An very important property of the boundary operator is that the composition operator $\partial_{k-1}\circ\partial_k$ is a zero map,
\begin{equation}
\begin{aligned}
\partial_{k-1}\partial_k(\sigma^k)&=\sum_{j<i}(-1)^i(-1)^j[u_0,...,\widehat{u_i},...\widehat{u_j},...u_k]+\sum_{j>i}(-1)^i(-1)^{j-1}[u_0,...,\widehat{u_j},...\widehat{u_i},...u_k] \\
&=0
\end{aligned}
\end{equation}
%With a sequence of homomorphisms of Abelian groups and $\partial_{k-1}\circ\partial_k=0$, we have a chain complex.
A sequence of chain groups connected by boundary operation form a chain complex,

\begin{equation}
\cdot\cdot\cdot \xrightarrow{\makebox[.27in]{}}C_n(\mathcal{K})\xrightarrow{\makebox[.27in]{$\partial_n$}}C_{n-1}(\mathcal{K})\xrightarrow{\makebox[.27in]{$\partial_{n-1}$}}\cdots\xrightarrow{\makebox[.27in]{$\partial_1$}}C_0(\mathcal{K})
\xrightarrow{\makebox[.27in]{$\partial_0$}}0.
\end{equation}
The equation $\partial_{k}\circ\partial_{k+1}=0$ is equivalent to the inclusion $Im\partial_{k+1} \in Ker \partial_{k}$, when $Im$ and $Ker$ denotes image and kernel. Elements of $Ker \partial_{k}$ are called $k$th cycle group, and denoted as $Z_k$=Ker$\partial_k$. Elements of $Im \partial_{k}$ are called  $k$th boundary group, and denoted as $B_k$=Im$\partial_{k+1}$.
%With cycle group $Z_k$ and boundary group $B_k$,
A $k$th homology group is defined as the quotient group of  $Z_k$ and  $B_k$.
\begin{equation}
H_k=Z_k/B_k.
\end{equation}
The $k$th Betti number of simplicial complex $\mathcal{K}$ is the rank of $H_k$, % corresponding to $\mathcal{K}$.
\begin{equation}
\beta_k=\text{rank}(H_k)=\text{rank}(Z_k)-\text{rank}(B_k).
\end{equation}
Betti number $\beta_k$ is finite number, since $\text{rank}(B_p)\leq\text{rank}(Z_p)<\infty$.
Betti numbers computed from a homology group are used to describe the corresponding space.
%For an object in $\mathbb{R}^3$,
Generally speaking, the Betti numbers $\beta_0$, $\beta_1$ and $\beta_2$ are numbers of connected components, tunnels, and cavities, respectively.

\noindent
\textbf{Filtration and persistence} A filtration of a simplicial complex $\mathcal{K}$ is a nested sequence of subcomplexes of $\mathcal{K}$.
\begin{equation}
\varnothing = \mathcal{K}_0 \subseteq \mathcal{K}_1 \subseteq ... \subseteq \mathcal{K}_m=\mathcal{K}.
\end{equation}
With a filtration of simplicial complex $\mathcal{K}$, topological attributes can be generated for each member in the sequence by deriving the homology group of each simplicial complex. The topological features that are long lasting through the filtration sequence are more likely to capture significant property of the object. Intuitively, non-boundary cycles that are not mapped into boundaries too fast along the filtration are considered to be possibly involved in major features or persistence. Equipped with a proper derivation of filtration and a wise choice of threshold to define persistence, it is practicable to filter out topological noises and acquire attributes of interest. The $p$-persistent $k$th homology group of $\mathcal{K}_i$ is defined as
\begin{equation}
H^{i,p}_k=Z^i_k/(B^{i+p}_k\cap Z^i_k),
\end{equation}
where $Z^i_k=Z_k(\mathcal{K}_i)$ and $B_k^i=B_k(\mathcal{K}_i)$. The consequent $p$-persistent $k$th Betti number is $\beta^{i,p}_k=\text{rank}(H^{i,p}_k)$. A well chosen $p$ promises reasonable elimination of topological noise. %Some practical methods to build abstract simplicial complexes from real world data and derivation of filtration are discussed in the following section.

\begin{figure}[H]
\centering
\subfigure[][]{\label{fig:ex1-a}
\includegraphics[width=.23\textwidth]{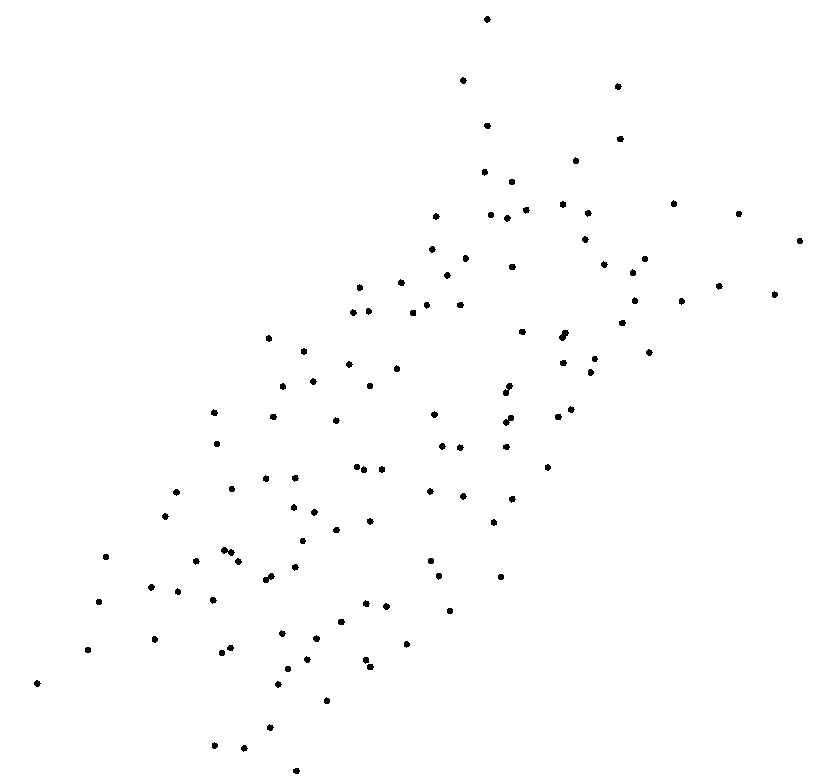}}
\subfigure[][]{\label{fig:ex1-b}
\includegraphics[width=.23\textwidth]{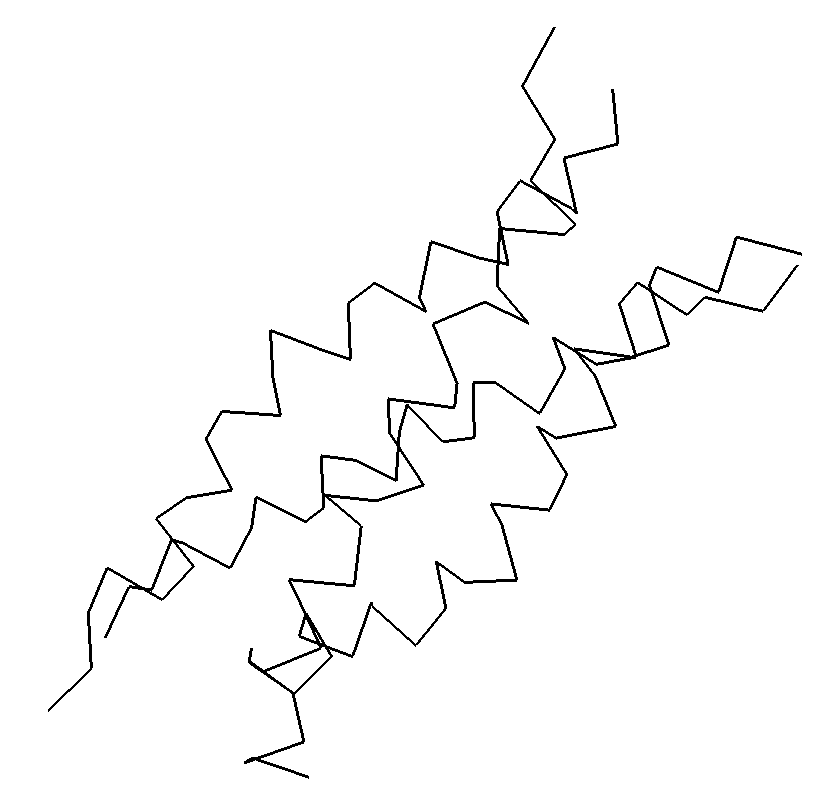}}
\subfigure[][]{\label{fig:ex1-c}
\includegraphics[width=.23\textwidth]{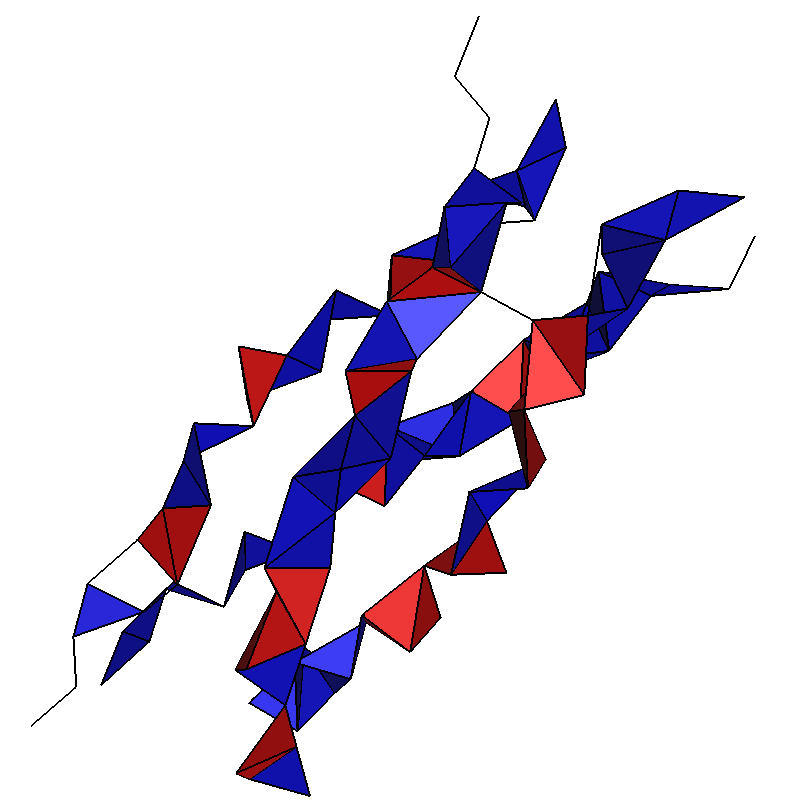}}
\subfigure[][]{\label{fig:ex1-d}
\includegraphics[width=.23\textwidth]{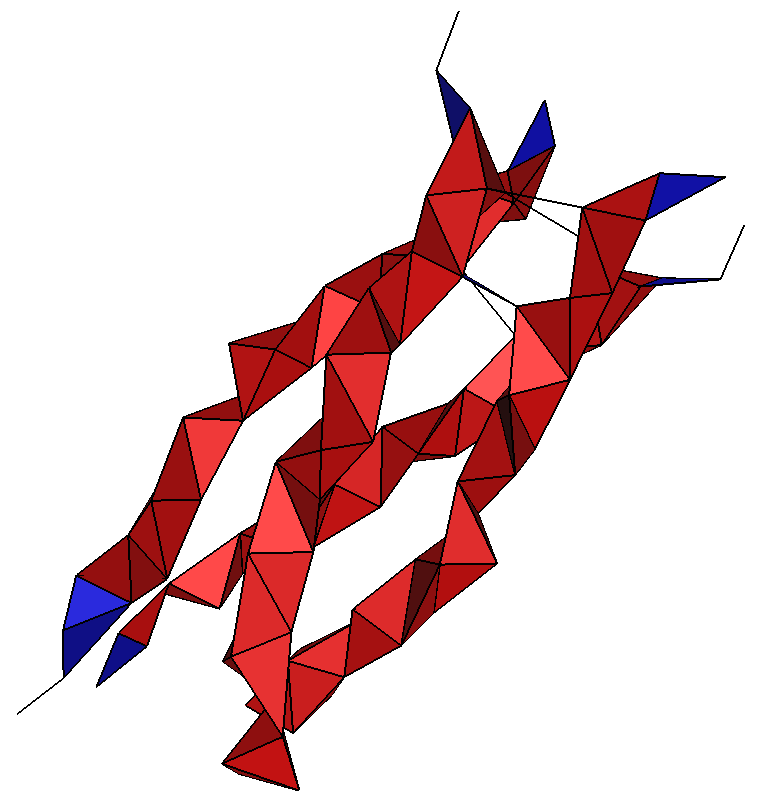}} \\
\subfigure[][]{\label{fig:ex1-e}
\includegraphics[width=.23\textwidth]{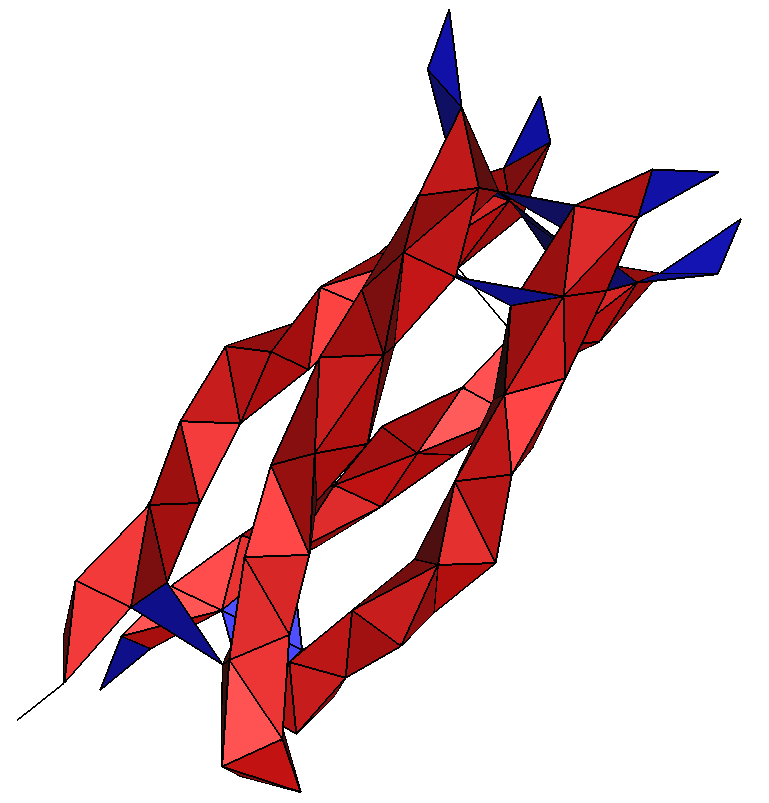}}
\subfigure[][]{\label{fig:ex1-f}
\includegraphics[width=.23\textwidth]{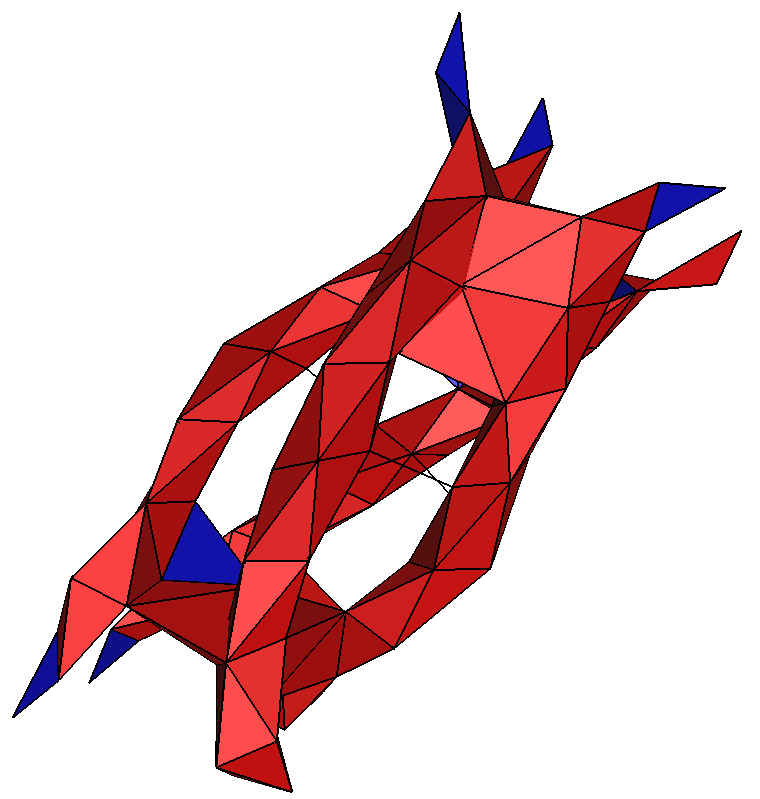}}
\subfigure[][]{\label{fig:ex1-g}
\includegraphics[width=.23\textwidth]{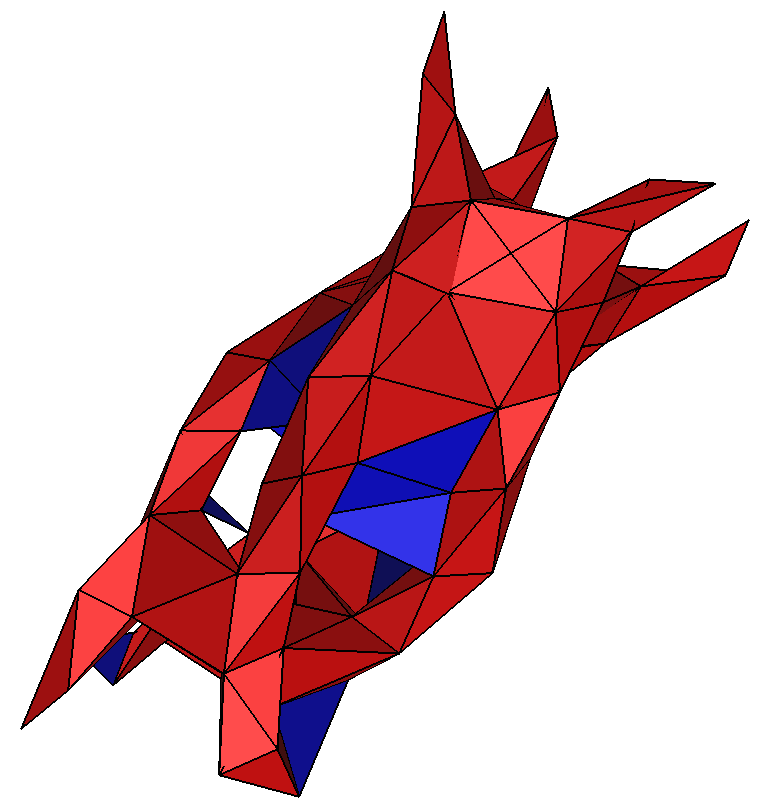}}
\subfigure[][]{\label{fig:ex1-h}
\includegraphics[width=.23\textwidth]{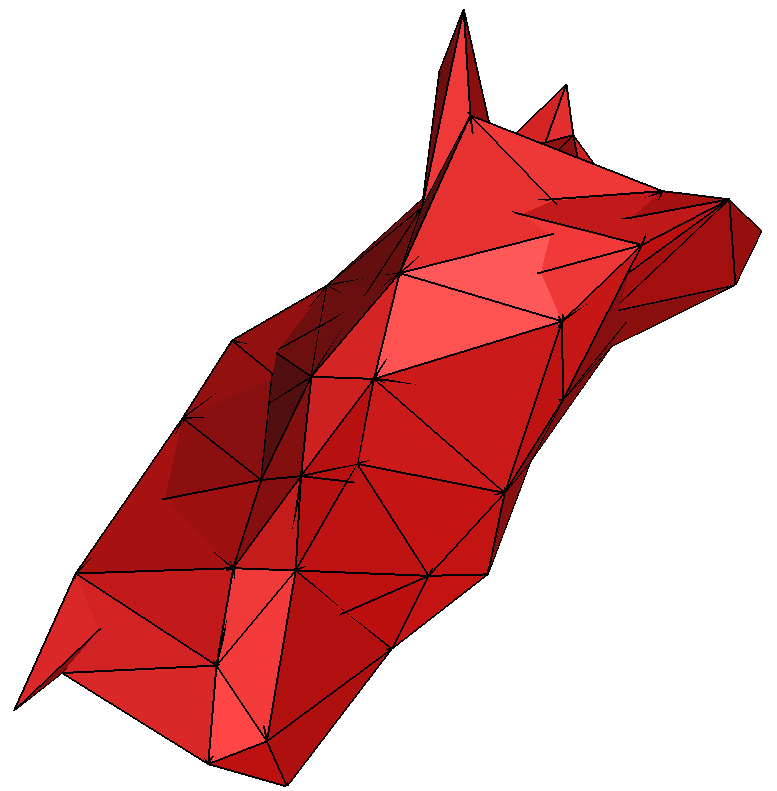}}
\caption{Filtration of Vietoris-Rips complex built on the $\alpha$-carbon point cloud of the M2 chimera channel of influenza A virus (PDB ID: 2LJC \cite{Pielak:2011}). The corresponding filtration values for each graph are \subref{fig:ex1-a} $d=1.0$\AA, \subref{fig:ex1-b} $d=4.0$\AA, \subref{fig:ex1-c} $d=5.5$ \AA, \subref{fig:ex1-d} $d=6.0$\AA, \subref{fig:ex1-e} $d=7.0$\AA, \subref{fig:ex1-f} $d=8.0$\AA, \subref{fig:ex1-g} $d=9.0$\AA, and \subref{fig:ex1-h} $d=12.0$\AA. Facets of $3$-simplices are shown in red, $2$-simplices are shown in blue, $1$-simplices are shown as lines, and $0$-simplices are shown as dots.}
\label{fig:ex1}
\end{figure}

\textbf{Vietoris-Rips complex} Based on a metric space $M$ and a given cutoff distance $d$, an abstract simplicial complex can be built. If two points in $M$ have a distance shorter than the given distance $d$, an edge is formed between these two points. Consequently, simplices of different dimensions are formed and a simplicial complex is built. For a point cloud data, natural metric space based on Euclidean distance or other metric spaces based on alternative definition of distance can be used to build a Vietoris-Rips complex. For example, any correlation matrix %like matrix from Gaussian network model of biomolecules
can be used directly to form a Vietoris-Rips complex. Figure~\ref{fig:ex1} illustrates growth of Vietoris-Rips complex along with increment of $d$ over the point set of $C_\alpha$ atoms from M2 chimera channel.

\begin{figure}[H]
\centering
\subfigure[][]{\label{fig:ex2-a}
\includegraphics[width=.3\textwidth]{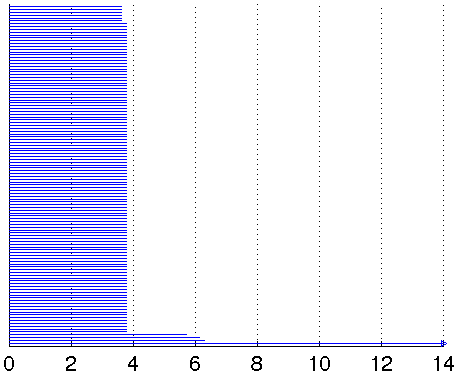}}
\subfigure[][]{\label{fig:ex2-b}
\includegraphics[width=.3\textwidth]{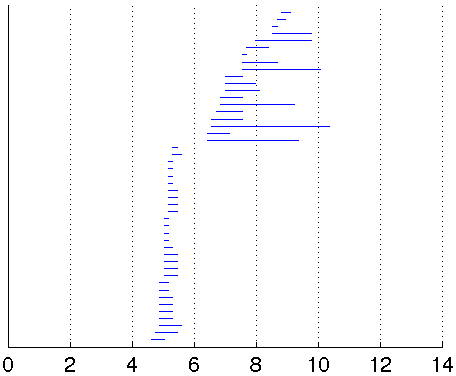}},   \\
\subfigure[][]{\label{fig:ex2-c}
\includegraphics[width=.3\textwidth]{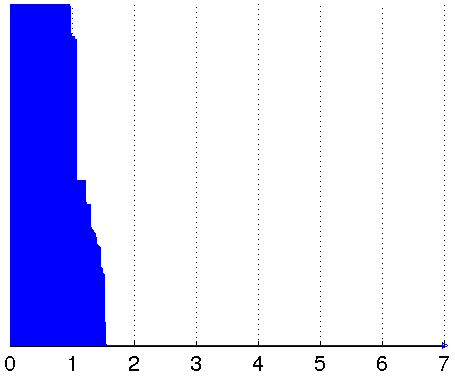}}
\subfigure[][]{\label{fig:ex2-d}
\includegraphics[width=.3\textwidth]{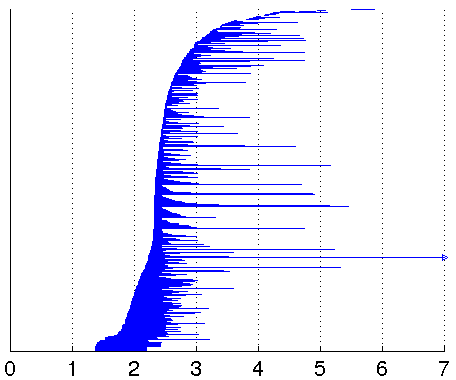}}
\caption{Bar code plots of persistent homology calculated for $\alpha$-carbon and all atom point cloud of M2 chimera channel of influenza A virus based in Vietoris-Rips complex. \subref{fig:ex2-a} and \subref{fig:ex2-b} are respectively Betti 0 and Betti 1 bar code plots for point cloud of $\alpha$-carbon. \subref{fig:ex2-c} and \subref{fig:ex2-d} are respectively Betti 0 and Betti 1 bar code plots for all atom point cloud.}
\label{fig:ex2}
\end{figure}

There are many ways of constructing complex other than Vietoris-Rips complex, including Alpha complex, Cech Complex, CW complex, etc. In the present work, we used Vietoris-Rips complex in part because of its intuitive nature and in part because of the moderate size of the systems we studied. The computational topology package JavaPlex\cite{Javaplex:2014} was used for computation of persistent homology. The results were represented in the form of barcodes \cite{Ghrist:2008}. %which is also known as topological fingerprints.
Figure~\ref{fig:ex2} illustrates barcodes computed from a point cloud data extracted from $C_{\alpha}$ atoms of protein ID 2LJC. % biomolecule.

\subsection{Support vector machine  }\label{svm+roc}

\paragraph{Basic theory}
SVM is a machine learning method that can be applied to classification and regression problems. It computes a hyperplane which maximizes margin between positive and negative training sets. In this work, Classification SVM Type 1, also known as C-support vector classification (C-SVC) \cite{Cortes:1995} is used. For the problem of classification, with pre-determined classes, a classifier is trained on a data set with the description of samples and their classes and it predicts the class of a new observation. The input for SVM is a set of samples. Each  sample  has a feature vector that describes the properties of the sample and a label that implies to which class the sample belongs. Given the input which is the training set, SVM will generate a hyperplane in the feature space or higher dimensional spaces depending on which kernel it uses that separates the classes. For two-class SVM, it looks for a hyperplane $\mathbf{w}^T\mathbf{x}+b=0$ that separates the classes. The determination of the coefficients $\mathbf{w}$ and $b$ breaks down to a constrained  optimization problem as
\begin{equation}\label{eq:opt-1}
\underset{\mathbf{w},b}{\text{min}}\frac{1}{2}|\mathbf{w}|^2+C\sum\xi_i,
\end{equation}
subject to
\begin{equation}
\begin{cases}
y_i(\mathbf{w}^T\mathbf{x}_i+b)\geq 1-\xi_i, i=1,...,n, \\
\xi_i\geq 0, i=1,...,n,
\end{cases}
\end{equation}
where $\mathbf{x}_i$ denotes the feature vector of the $i$th sample, $y_i$ is the label of the $i$th sample which takes value of either $1$ or $-1$, and  $C$ is a penalty coefficient for misclassified points. To handle linearly inseparable data, one maps the data into a higher dimensional space as $\phi:\mathbb{R}^N\rightarrow\mathbb{R}^M$ with $N<M$. Since in the optimization problem and in scoring function of the classifier, the operator used is dot product, $\phi$ does not need to be explicitly found. A decaying kernel $K(\mathbf{x}_i,\mathbf{x}_j)$ function is used to represent $\phi^T(\mathbf{x}_i)\phi(\mathbf{x}_j)$. Commonly used kernel functions include linear function: $K(\mathbf{x}_i,\mathbf{x}_j)=\mathbf{x}_i^T\mathbf{x}_j$, polynomial: $K(\mathbf{x}_i,\mathbf{x}_j)=(a\mathbf{x}_i^T\mathbf{x}_j+b)^d$, radial basis functions (RBFs) such as Gaussian  $K(\mathbf{x}_i,\mathbf{x}_j)=\exp(-\gamma|\mathbf{x}_i-\mathbf{x}_j|^2),\gamma>0$. In fact, the admissible kernels of fleibility-rigidity index (FRI) \cite{KLXia:2013d,Opron:2014,Opron:2015a} work too.  In this work, The Gaussian kernel is used and a 5-fold cross validation was applied to search for optimized training parameters for problems with large amount of samples. To solve the optimization problem, the original problem is transformed into the corresponding Lagrange dual problem. For a contained optimization problem
\begin{equation}\label{LDP-1}
\begin{split}
\underset{\mathbf{x}}{\text{min}}f(\mathbf{x}), \\ %\mathbf{x}\in\Omega \\
g_i(\mathbf{x})\leq 0, i=1,2,...,k_1 \\
h_i(\mathbf{x})=0, i=1,2,...,k_2
\end{split}
\end{equation}
the Lagrange function of this problem is defined as
\begin{equation}
L(\mathbf{x},\boldsymbol{\alpha},\boldsymbol{\lambda})=f(\mathbf{x})+\sum_{i=1}^{k_1}\alpha_ig_i(\mathbf{x})+\sum_{i=1}^{k_2
}\lambda_ih_i(\mathbf{x})
\end{equation}
where $\boldsymbol{\alpha}$ and $\boldsymbol{\lambda}$ are Lagrange multipliers.
The Lagrange dual problem is defined as
\begin{equation}
\begin{split}
\underset{\boldsymbol{\alpha},\boldsymbol{\lambda}}{\text{max}}\theta(\boldsymbol{\alpha},\boldsymbol{\lambda}) \\
\alpha_i\geq 0
\end{split}
\end{equation}
where $\theta(\boldsymbol{\alpha},\boldsymbol{\lambda})=\underset{\mathbf{x}\in\Omega}{\text{inf}}L(\mathbf{x},\boldsymbol{\alpha},\boldsymbol{\lambda})$.
The Lagrange function of the original optimization problem (\ref{eq:opt-1}) is formulated as
\begin{equation}\label{LDP-2}
\begin{aligned}
L(\mathbf{w},b,\boldsymbol{\xi},\boldsymbol{\alpha},\boldsymbol{\lambda}) &= \frac{1}{2}\mathbf{w}^T\mathbf{w}+\sum_{i=1}^n(C-\alpha_i-\lambda_i)\xi_i-b\sum_{i=1}^n\alpha_iy_i+\sum_{i=1}^n\alpha_iy_i\mathbf{w}^T\phi(\mathbf{x}_i) \\
&= \sum_{i=1}^n\alpha_i-\frac{1}{2}\sum_{i,j=1}^ny_iy_j\alpha_i\alpha_j\phi(\mathbf{x}_i)^T\phi(\mathbf{x}_j).
\end{aligned}
\end{equation}
Tthe corresponding dual problem with Karush-Kuhn-Tucker conditions is defined as
\begin{equation}\label{eq:final-dual}
\begin{split}
\underset{\boldsymbol{\alpha}}{\text{max}} L(\mathbf{w},b,\boldsymbol{\xi},\boldsymbol{\alpha},\boldsymbol{\lambda}) \\
\alpha_i(y_i(\sum_{j=1}^n\alpha_j\phi(\mathbf{x}_j)^T\phi(\mathbf{x}_i)+b)-1+\xi_i))=0 \\
\xi_i(\alpha_i-C)=0 \\
\sum_{i=1}^n\alpha_iy_i=0 \\
C\geq\alpha_i\geq 0
\end{split}
\end{equation}
The dual problem can be solved with sequential minimal optimization (SMO) method \cite{Fan:2005}.

\paragraph{Receiver operating characteristic (ROC)}

ROC is a plot that visualizes the  performance of a binary classifier \cite{Fawcett:2006}. A binary classifier uses a threshold value to decide the prediction label of an entry. In testing process, we define true positive rate (TPR) and false positive rate (FPR) for the testing set.
\begin{equation}\label{TPR/FPR}
\begin{aligned}
\text{TPR} &= (\text{number of positive samples predicted as positive})/(\text{number of positive samples}) \\
\text{FPR} &= (\text{number of nagetive samples predicted as positive})/(\text{number of nagetive samples})
\end{aligned}
\end{equation}
An ROC space is a two dimensional space defined by points with $x$ coordinate representing FPR and $y$ coordinate representing TPR. In the prediction process of a binary classifier, a score is assigned to a sample by the classifier. A test sample may be labeled as positive or negative with different threshold value used by the classifier. Corresponding to a certain threshold value, there is a pair of FPR and TPR values which is a point in the ROC space. All such points will fall in the box $[0,1]\times[0,1]$. Points above the diagonal line $y=x$ are considered as  good predictors and those below the line are considered as poor predictors. If a point is below the diagonal line, the predictor can be inverted to be a good predictor. For points that are close to the diagonal line, they are considered to act similarly to random guess which implies a relatively useless predictor. ROC curve is obtained by plotting FPR and TPR as continuous functions of threshold value. The area between ROC curve and $x$ axis represents probability that the classifier assigns higher score to a randomly chosen positive sample than to a randomly chosen negative sample if positive is set to have higher score than negative. The area under the  curve (AUC) of ROC is a measure of classifier quality. Intuitively, a higher AUC implies a better classifier.

\subsection{Topological feature selection and construction}\label{feature+preprocessing}

In this work, algebraic topology  is employed to discriminate proteins.  Specifically, we compute MTFs through the filtration process of protein structural data.  MTFs bear the persistence of  topological invariants during the filtration and are ideally suited for protein classification. To implement our topological approach in the SVM algorithm, we construct protein feature vectors by using MTFs.  We select distinguishing protein features from MTFs. These features can be both long lasting and short lasting Betti 0, Betti 1, and Betti 2 intervals.

Table~\ref{tab:features} lists  topological features used for classification. Detailed explanation of these features is discussed. The length and location value of bars are in the unit of  angstrom (\AA) for protein data.
\begin{table}[ht]
\centering
\caption{A list of features used in support vectors}
\begin{tabular}{|c|c|l|}
\hline
Feature \# & Betti \# & Description \\
\hline
 1 & 0 & The length of the second longest Betti $0$ bar.\\
 2 & 0 & The length of the third longest Betti $0$ bar. \\
 3 & 0 & The summation of lengths of all Betti $0$ bars except for those exceed the max filtration value. \\
 4 & 0 & The average length of Betti $0$ bars  except for those exceed the max filtration value. \\
 5 & 1 & The onset value of the longest Betti $1$ bar. \\
 6 & 1 & The length of the longest Betti $1$ bar. \\
 7 & 1 & The smallest onset value of the Betti $1$ bar that is longer than  1.5\AA. \\
 8 & 1 & The average of the  middle point values  of all the Betti $1$ bars that are  longer than  1.5\AA. \\
 9 & 1 & The number of Betti $1$ bars that   locate at   $[4.5, 5.5]$\AA,  divided by the number of atoms.\\
 10 & 1 & The number of Betti $1$ bars that locate  at $[3.5, 4.5)$\AA~ and   $(5.5,6.5]$\AA,   divided by the number of atoms. \\
 11 & 1 & The summation of lengths of all the Betti $1$ bars except for those exceed the max filtration value. \\
 12 & 1 & The average length of Betti $1$ bars  except for those exceed the max filtration value.\\
 13 & 2 & The onset value of the first Betti $2$ bar that ends after a given number. \\
\hline
\end{tabular}
\label{tab:features}
\end{table}

\begin{itemize}
\item Feature 1: The length of the second longest Betti $0$ bar indicates the onset in  filtration that the simplices in the corresponding complex form one connected component.
\item Feature 2: Similar to Feature 1, this value indicates the onset in filtration that the simplices form two connected components. For the more complicated point cloud, the more features of this kind may be utilized.
\item Feature 3: Geometrically, the total length of Betti $0$ bars describes how compactly the points are located.
\item Feature 4: This averaged Betti $0$ bar length shows similar property as that in Feature 3 with no correlation to atom number.
\item Feature 5: This value shows the filtration value at which, the largest persistent loop is formed.
\item Feature 6: The persistence of the longest Betti $1$ bar reflects the size of the geometrically dominating loop.
\item Feature 7: A Betti $1$ bar with length larger than the threshold is considered to be important and this feature records the onset filtration value of such a long bar. In this work, a threshold of 1.5 is used for $\alpha$-carbon point cloud data of proteins.
\item Feature 8: This feature records the average location of midpoints of Betti $1$ bars which are longer than the threshold value discussed in Feature 5. This value shows at which filtration value the loops are centered.
\item Feature 9: This feature indicates the portion of alpha helices in a protein. For each four $\alpha$-carbons on a alpha helix, they are likely to form a short Betti $1$ bar around filtration value 5\AA. A bar is considered to be short if it has length shorter than 0.5\AA~ and to be around 5\AA~ if the distance from its midpoint to 5\AA~ is less than 0.6\AA.
\item Feature 10: Similar to Feature 7, this feature can be used to identify portion of beta sheets. Detailed discussion of Features 7 and 8 can be found in Ref. \cite{KLXia:2014c}.
\item Feature 11: A strong correlation between accumulation bar length of Betti $1$ and total energy has been reported \cite{KLXia:2014c}.
\item Feature 12: The average value of Betti $1$ bars correlates to the average loop size.
\item Feature 13: The smallest onset value of the Betti $2$ bar that ends after a given value. This feature gives information about birth and death of cavities in the complex through filtration.

\end{itemize}
Each feature was scaled to the interval $[0,1]$ with linear mapping.

\section{Results}\label{sec:Numerical}

In this section, we validate the proposed idea, examine the accuracy, explore the utility of the proposed topology based classification and analysis of protein molecules.  We consider four different types of problems. In our first case, we study a protein-drug binding problem, namely,   the drug inhibition of Influenza A virus M2 channels.  In our second case, we use MTFs to classify two type of conformations of hemoglobin proteins. Default parameters were used and brute force cross validation was performed for these first two cases due to their relatively small size of samples. We further consider the classification of three types of protein domains, i.e.,  all alpha domains, all beta domains and mixed alpha and beta domains. Finally,  our method was tested on a problem set, PCB00019, from Protein Classification Benchmark Collection \cite{Sonego:2007}. In the last two cases, a grid search with cross validation on training sets was performed to optimize SVM parameters. For the last case, different penalty parameters were applied to overcome the unbalanced data and an ROC analysis was used to evaluate the results.

Data for the M2 channels are all obtained from NMR experiments \cite{Pielak:2011}. Data for  hemoglobin structures are all collected from X-ray crystallography. Structure data for the last two test cases are mostly attained from X-ray crystallography. However, a few structures were determined by NMR techniques and thus have many alternatives. In this situation, we select the second structure for each sample in the data base.

In this work, we utilize JavaPlex \cite{Javaplex:2014} to compute MTFs.  For implementation of support vector machine, LIBSVM is employed  \cite{Chang:2011}.

\subsection{Protein-drug binding analysis}\label{sec:Case1}

%Two structures of M2 channel with and without rimantadine inhibition
%Add words about protein drug binding
  Proteins are vital to many processes in cells. In many biologic processes, protein may bind to other molecules. Protein-protein interaction and  protein-ligand interaction are of crucial importance to their functions and/or malfunction. These interactions have been intensively exploited in drug design. Specifically, many drugs bind to target proteins to modify their functions and activities. After binding to other molecule, a protein usually experiences a structure change at the binding site. In many cases, it may also undergo allosteric process with a global structural change upon the binding. We test our method for distinguishing proteins with drug bound from proteins without drugs.

We use  M2 channel, which is a transmembrane protein found in influenza A  virus \cite{Pielak:2011b}, as an example.   M2 channel equilibrates pH across the viral membrane during cell entry and palys a  vital role in the viral replication. Therefore, it is used  a target for the anti-influenza drugs, i.e., amantadine and rimantadine, which bind to the M2 channel pore and thus block the proton permeation.  The drug binding creates a topological change to the M2 channel in the conventional sense. However, in the present work,  it is not the topological change itself, rather that the binding induced geometric variation of the M2 channel that is converted into the change in the topological invariants. Such a change is recorded in our MTFs and utilized for protein classification.

\begin{figure}[H]
\centering
\subfigure[][]{\label{fig:ex3-a}
\includegraphics[width=.4\textwidth]{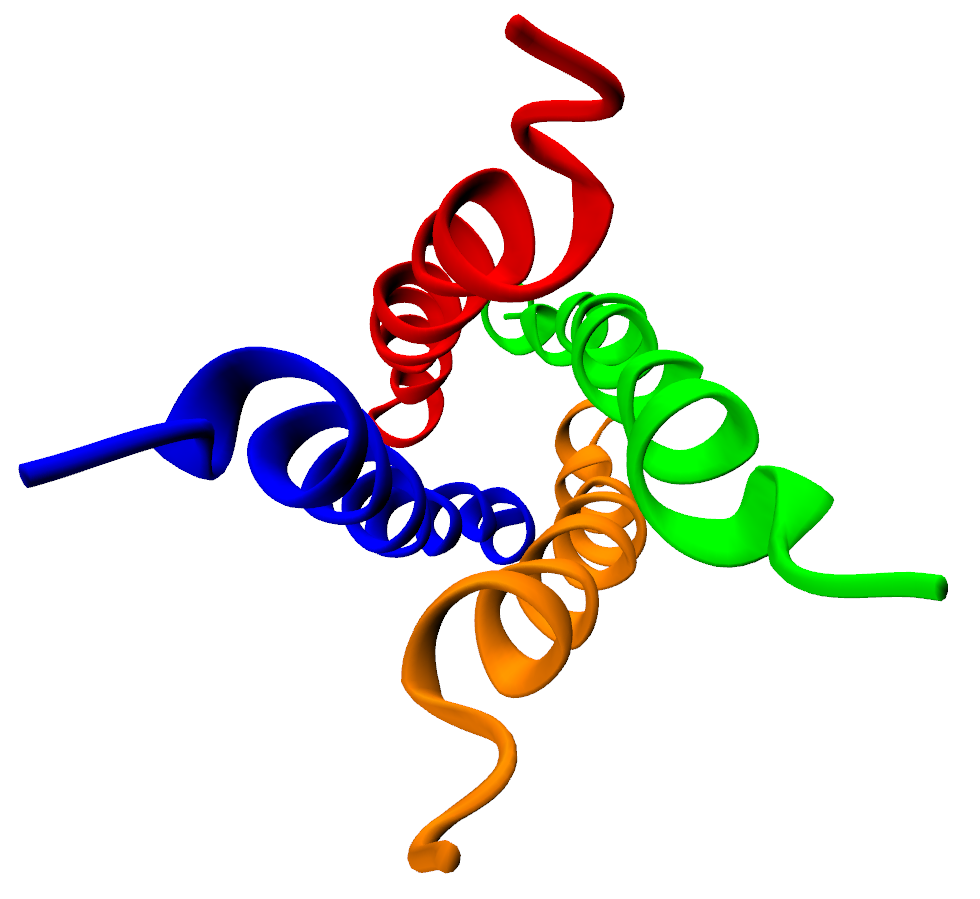}}
\subfigure[][]{\label{fig:ex3-b}
\includegraphics[width=.4\textwidth]{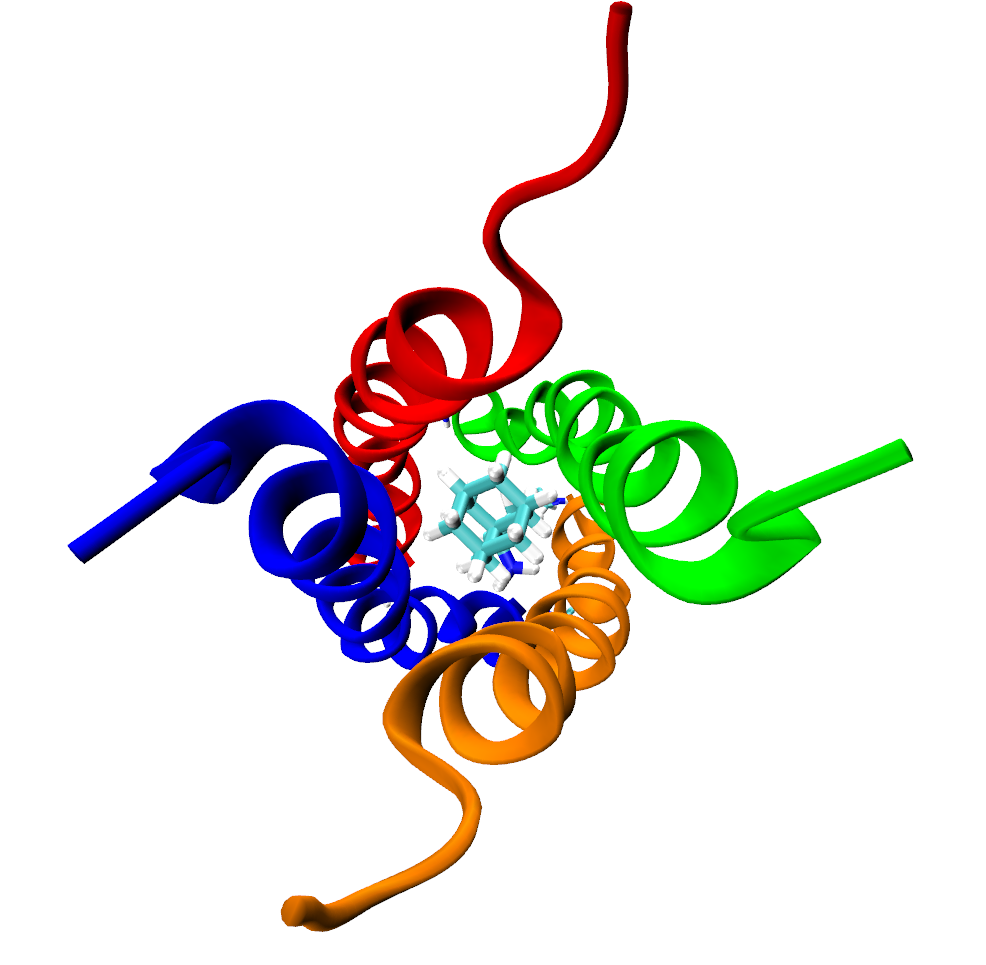}} \\
\subfigure[][]{\label{fig:ex3-1}
\includegraphics[width=.29\textwidth]{M2-B-betti0-new1.png}}
\subfigure[][]{\label{fig:ex3-2}
\includegraphics[width=.29\textwidth]{M2-B-betti1-new1.png}}
\subfigure[][]{\label{fig:ex3-3}
\includegraphics[width=.29\textwidth]{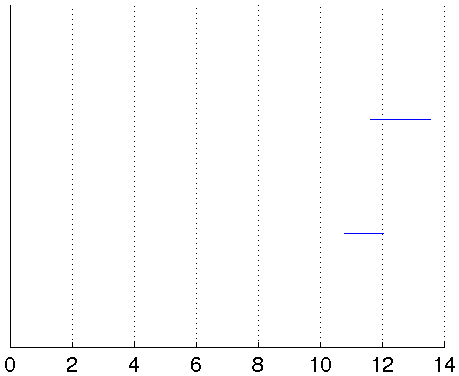}} \\
\subfigure[][]{\label{fig:ex3-7}
\includegraphics[width=.29\textwidth]{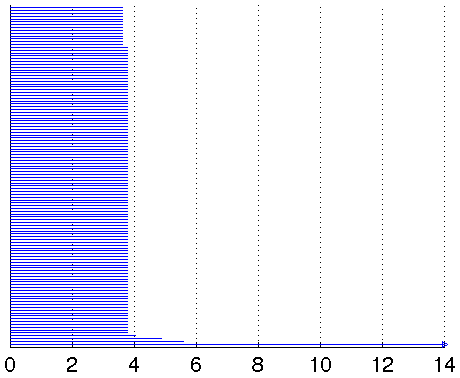}}
\subfigure[][]{\label{fig:ex3-8}
\includegraphics[width=.29\textwidth]{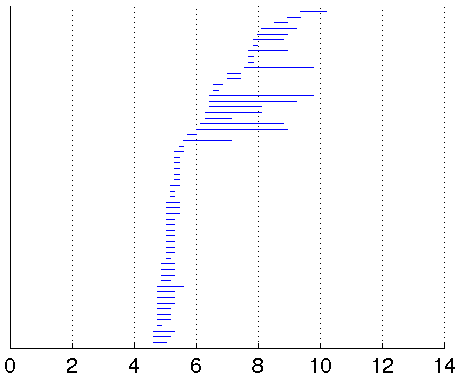}}
\subfigure[][]{\label{fig:ex3-9}
\includegraphics[width=.29\textwidth]{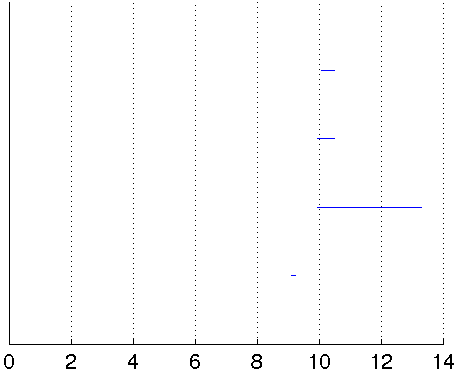}} \\
\caption{Protein structures used in  M2 channel classification. \subref{fig:ex3-a} (PDB ID: 2LJB \cite{Pielak:2011}) M2 channel of influenza A without inhibitor. \subref{fig:ex3-b} (PDB ID: 2LJC\cite{Pielak:2011}) M2 channel of influenza A with inhibitor. The small molecule in the graph is for illustration and was not used in classification. \subref{fig:ex3-1}, \subref{fig:ex3-2} and \subref{fig:ex3-3} are respectively Betti 0, Betti 1, and Betti 2 barcodes for \subref{fig:ex3-a}. \subref{fig:ex3-7}, \subref{fig:ex3-8} and \subref{fig:ex3-9} are respectively Betti 0, Betti 1, and Betti 2 barcodes for \subref{fig:ex3-b}.}
\label{fig:ex3}
\end{figure}

The structures of chimera channels with and without rimantadine were used for classification. PDB IDs of the two structures are 2LJC for channel with the inhibitor and 2LJB for channel without the inhibitor \cite{Pielak:2011}.  The structures are shown in Figure~\ref{fig:ex3-a}--\subref{fig:ex3-b}. Note that inhibit itself is not included in our filtration. A total of 15 snapshots from NMR for each structure is used to perform classification. Due to small size of instances, default parameter in C-SVC with penalty $C=2$ and $\gamma = 1/(\text{number of features})$ were used. Each time, 10 instances from each class were set as the training set and the rest were set as the testing set. A brute-force cross validation was performed. The average accuracy for unbound form is 93.91\% and accuracy for bound form is 98.31\%. Due to small size of testing set, AUC value was not calculated in this example.

\subsection{Discrimination of hemoglobin molecules in relaxed and taut forms}\label{sec:Case2}

\begin{figure}[H]
\centering
\subfigure[][]{\label{fig:ex3-c}
\includegraphics[width=.4\textwidth]{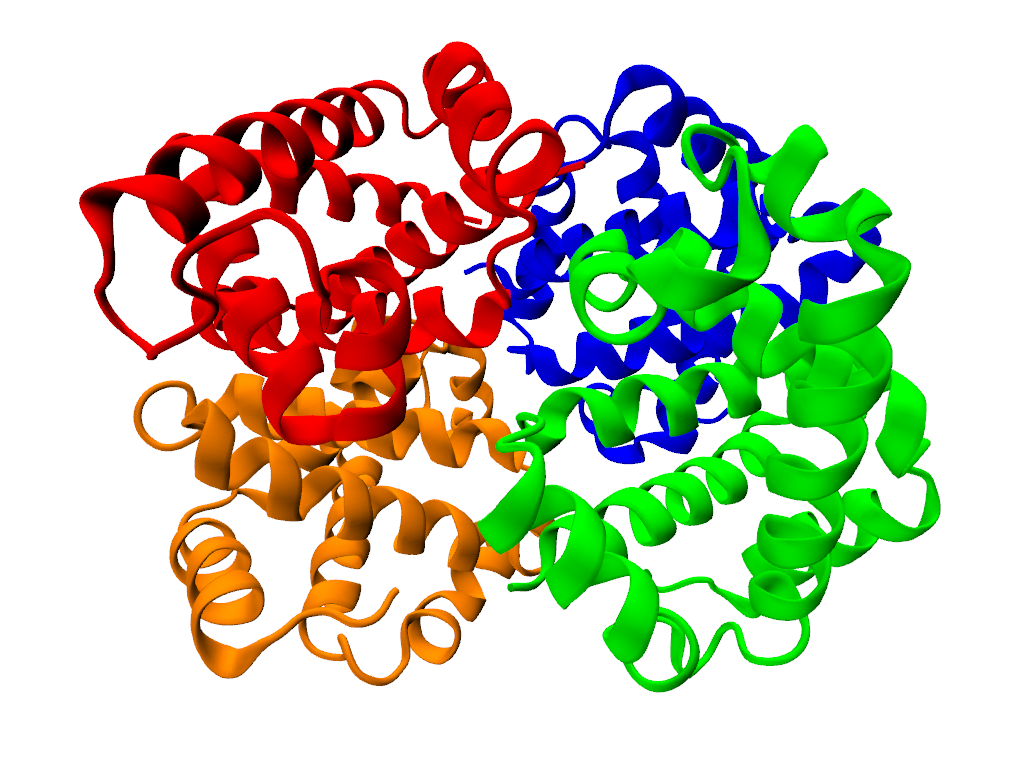}}
\subfigure[][]{\label{fig:ex3-d}
\includegraphics[width=.4\textwidth]{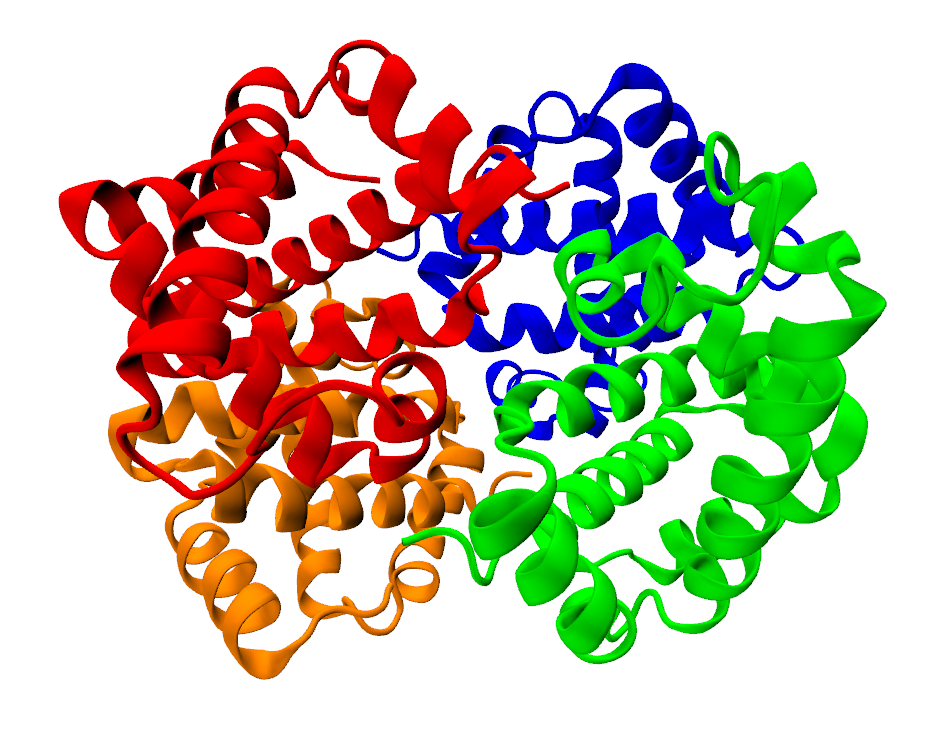}} \\
\subfigure[][]{\label{fig:ex3-4}
\includegraphics[width=.29\textwidth]{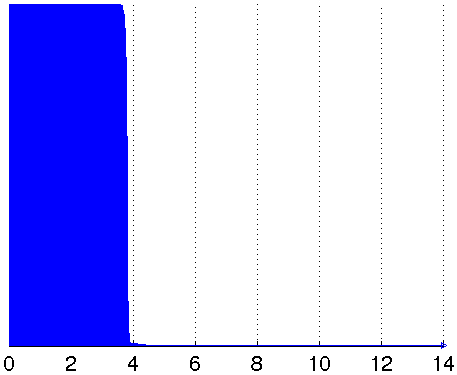}}
\subfigure[][]{\label{fig:ex3-5}
\includegraphics[width=.29\textwidth]{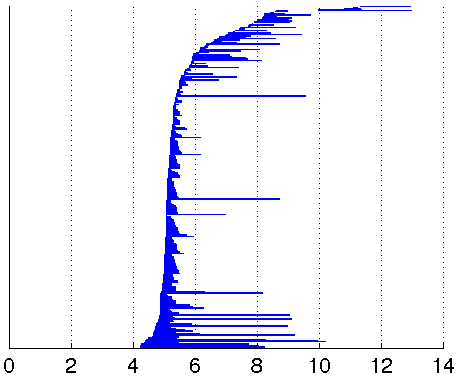}}
\subfigure[][]{\label{fig:ex3-6}
\includegraphics[width=.29\textwidth]{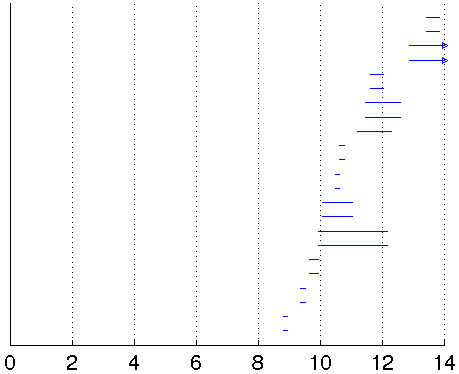}} \\
\subfigure[][]{\label{fig:ex3-10}
\includegraphics[width=.29\textwidth]{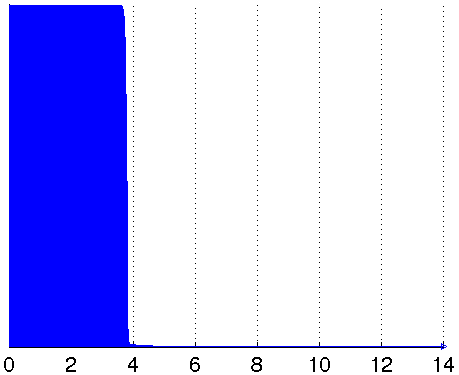}}
\subfigure[][]{\label{fig:ex3-11}
\includegraphics[width=.29\textwidth]{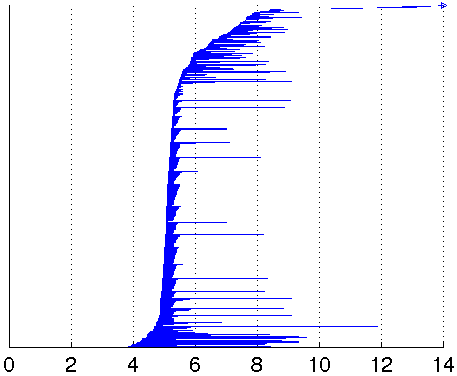}}
\subfigure[][]{\label{fig:ex3-12}
\includegraphics[width=.29\textwidth]{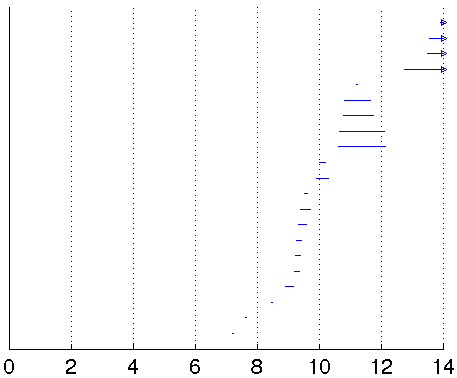}}
\caption{Protein structures used  the Hemoglobin classification. \subref{fig:ex3-c} (PDB ID: 3A0G) Relaxed (R) form of hemoglobin which express high affinity to oxygen. \subref{fig:ex3-d} (PDB ID: 2HHB\ cite{Fermi:1984}) Taut (T) form of hemoglobin which express low affinity to oxygen. \subref{fig:ex3-4}, \subref{fig:ex3-5} and \subref{fig:ex3-6} are respectively Betti 0, Betti 1, and Betti 2 barcodes for \subref{fig:ex3-c}. \subref{fig:ex3-10},  \subref{fig:ex3-11} and \subref{fig:ex3-12} are respectively  Betti 0, Betti 1, and Betti 2 barcodes for \subref{fig:ex3-d}.}
\label{fig:ex32}
\end{figure}

Hemoglobin is oxygen transport metallprotein in red blood cells of most vertebrates. It carries oxygen from lungs or gills to other organs  or parts in the body. Oxygen is released to tissues and is used for metabolism. Hemoglobin  is also known to carry carbon dioxide in some cases. It exits in two forms, known as taut (T) form and relaxed (R) form. Examples of  these two forms are shown in Figure~\ref{fig:ex3-c}--\subref{fig:ex3-d}. Relaxed form has a high oxygen binding affinity with which hemoglobin can better bind to oxygen in lungs or gills. Taut form has a low oxygen binding affinity which helps release the oxygen in the rest of the body. Many factors affect the conformation form of hemoglobin, such as pH value, concentration of carbon dioxide and partial pressure in the system. Structurally, the two forms are slightly different. In this test case, we picked 9 structures of hemoglobin in R form and 10 structures of hemoglobin in T form from protein data bank. Table \ref{tab:hemo-PDB}  lists of PDB IDs used.
\begin{table}[ht]
\centering
\caption{Protein molecules used for the Hemoglobin classification. }
\begin{tabular}{|c|c|}
\hline
R-form & 1HHO, 3A0G, 1LFQ, 1HBR, 1RVW, 2D5X, 1IBE, 1AJ9, 2W6V \\
\hline
T-form & 2HHB, 2DHB, 1LFL, 2D5Z, 1GZX, 2HBS, 4ROL, 1O1J, 2DXM, 1KD2\\
\hline
\end{tabular}
\label{tab:hemo-PDB}
\end{table}
\noindent

In this test case, as the number of instances is relatively small, a brute-force cross validation was performed with the same default parameters as in last case.  Each time one instance from each class were picked as the test set leaving the rest instances as the training set. The average accuracy of the prediction for test set is 84.50\%. The average accuracy of R form is 77.16\% and average accuracy of T form is 91.11\%. Since test set size is small, ROC analysis was not applied in this case.

\subsection{The classification of all alpha, all beta, and mixed alpha and beta protein domains}\label{sec:Case3}

Protein secondary structures are three dimensional patterns of protein local segments. Common secondary structures include alpha helices and beta sheets. These local structures are formed by   hydrogen bonds between amine hydrogen and carbonyl oxygen atoms in the backbone of a protein. Typically, secondary structures can  be identified from amino acid sequence data. In this test example, we use only geometric data without sequence information to generate MTFs and then classify alpha helices and beta sheets. Instances of this example were taken from \href{http://scop.berkeley.edu/}{ SCOPe} (Structural Classification of Proteins-extended) database \cite{Fox:2014}. The SCOPe ID (SID) of samples used in this test case  are listed in Tables \ref{tab:allalpha},\ref{tab:allbeta}, and \ref{tab:aplb}.

\begin{figure}[H]
\centering
\subfigure[][]{\label{fig:ex4-a}
\includegraphics[width=.3\textwidth]{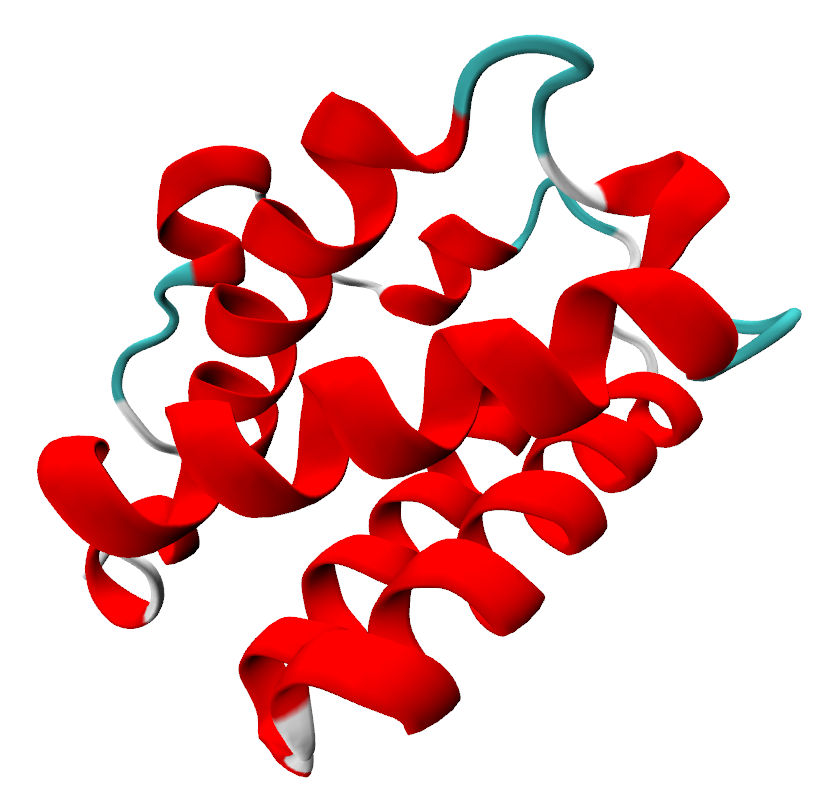}}
\subfigure[][]{\label{fig:ex4-b}
\includegraphics[width=.3\textwidth]{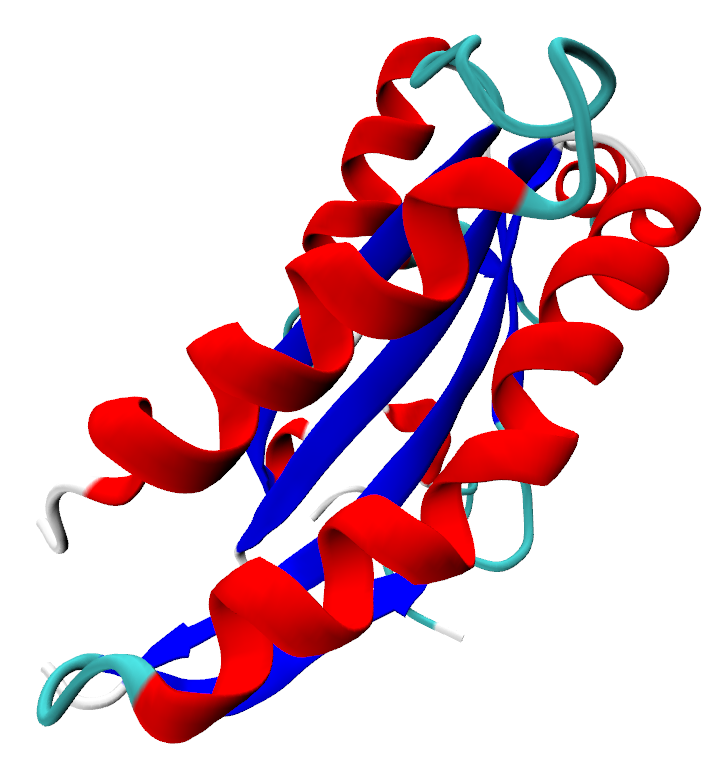}}
\subfigure[][]{\label{fig:ex4-c}
\includegraphics[width=.3\textwidth]{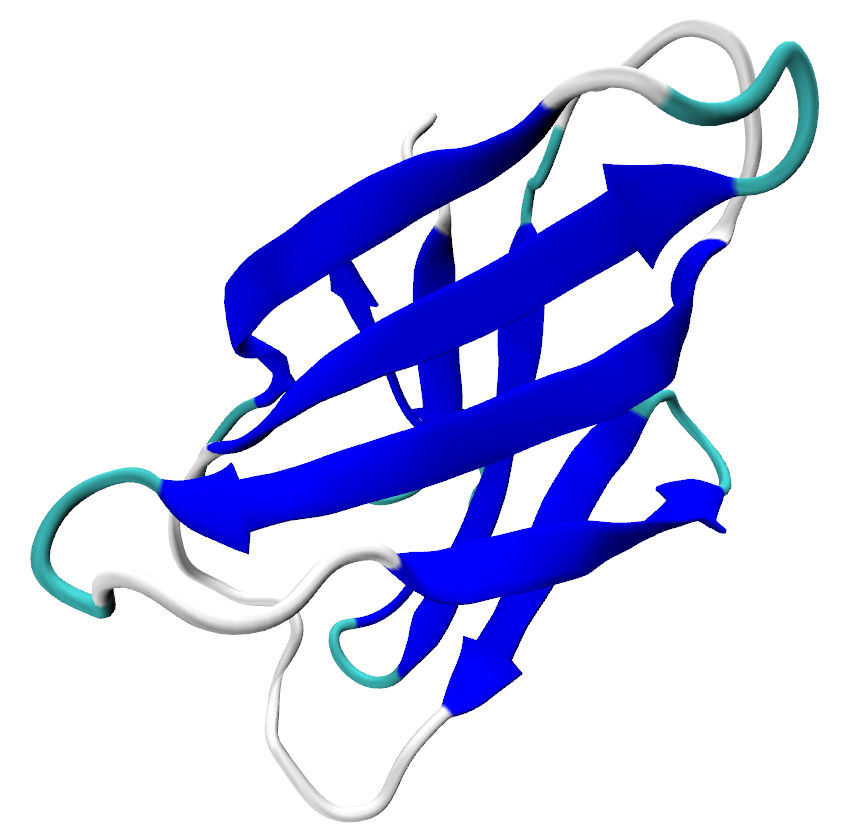}} \\
\subfigure[][]{\label{fig:ex4-1}
\includegraphics[width=.3\textwidth]{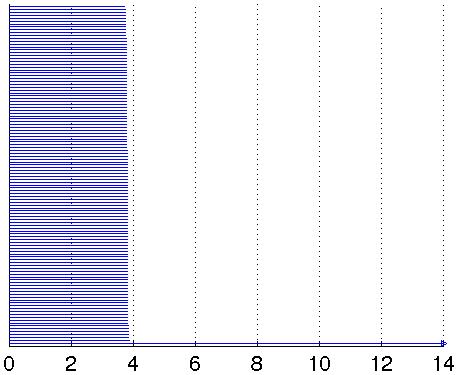}}
\subfigure[][]{\label{fig:ex4-2}
\includegraphics[width=.3\textwidth]{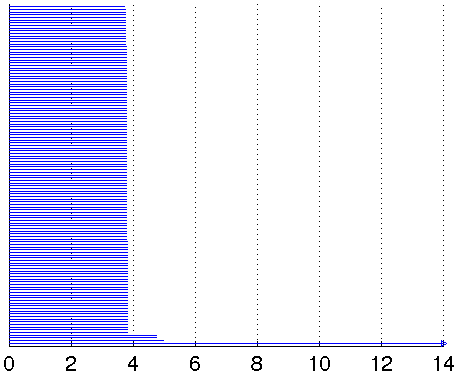}}
\subfigure[][]{\label{fig:ex4-3}
\includegraphics[width=.3\textwidth]{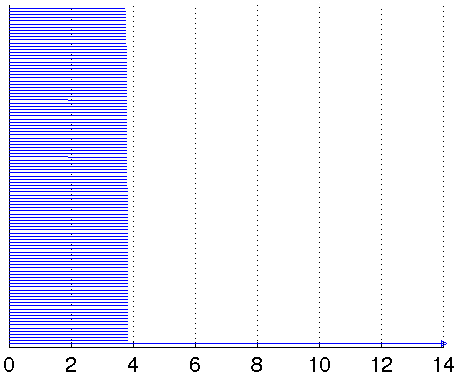}} \\
\subfigure[][]{\label{fig:ex4-4}
\includegraphics[width=.3\textwidth]{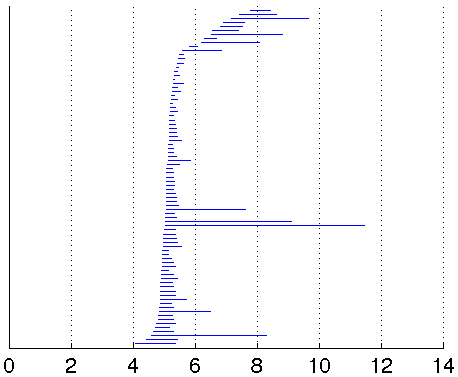}}
\subfigure[][]{\label{fig:ex4-5}
\includegraphics[width=.3\textwidth]{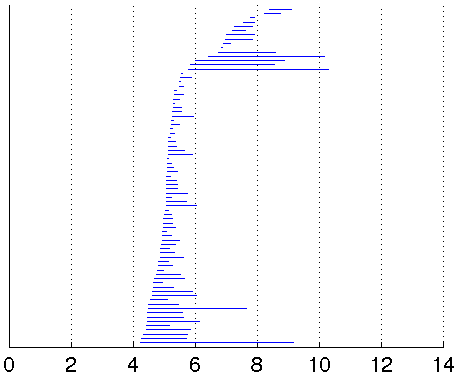}}
\subfigure[][]{\label{fig:ex4-6}
\includegraphics[width=.3\textwidth]{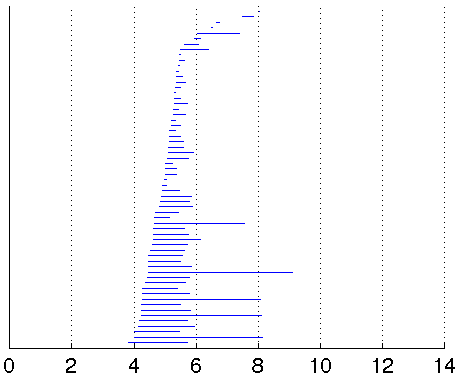}}
\caption{ Example plots  of different protein domains. \subref{fig:ex4-a} All alpha protein. \subref{fig:ex4-b} Alpha and beta protein. \subref{fig:ex4-c} All beta protein. \subref{fig:ex4-1} and \subref{fig:ex4-4} are respectively example Betti 0 and Betti 1 barcodes for all alpha protein. \subref{fig:ex4-2} and \subref{fig:ex4-5} are respectively example Betti 0 and Betti 1 barcodes for alpha+beta protein. \subref{fig:ex4-3} and \subref{fig:ex4-6} are respectively  example Betti 0 and Betti 1 barcodes for all beta protein.
}
\label{fig:ex4}
\end{figure}

In this test case, protein domains were separated into three classes, namely, all alpha helix domains, all beta domains, and mixed alpha and beta domains. Examples for each of three   classes are shown in Figures~\ref{fig:ex4-a}--\subref{fig:ex4-c} and their barcode plots are shown in Figures~\ref{fig:ex4-1}--\subref{fig:ex4-6}. For each class in SCOPe, 300  structures from different superfamilies were used for classification. Among the 900 instances, 60 from each class were used as the test set and the rest were used as the training set. A 5-fold cross-validation was performed to test accuracy. In each training process, a 5-fold cross validation in the training set was carried out to optimize training parameters. The overall accuracy is 84.93\%. Specifically, the accuracy for all alpha helix domains is 90.67\%, the accuracy for mixed alpha and beta domains is 78.77\%, and accuracy for all beta domains is 83.31\%.

\subsection{Classification of protein superfamilies}\label{sec:Case4}

\begin{figure}[H]
\centering
\includegraphics[width=.6\textwidth]{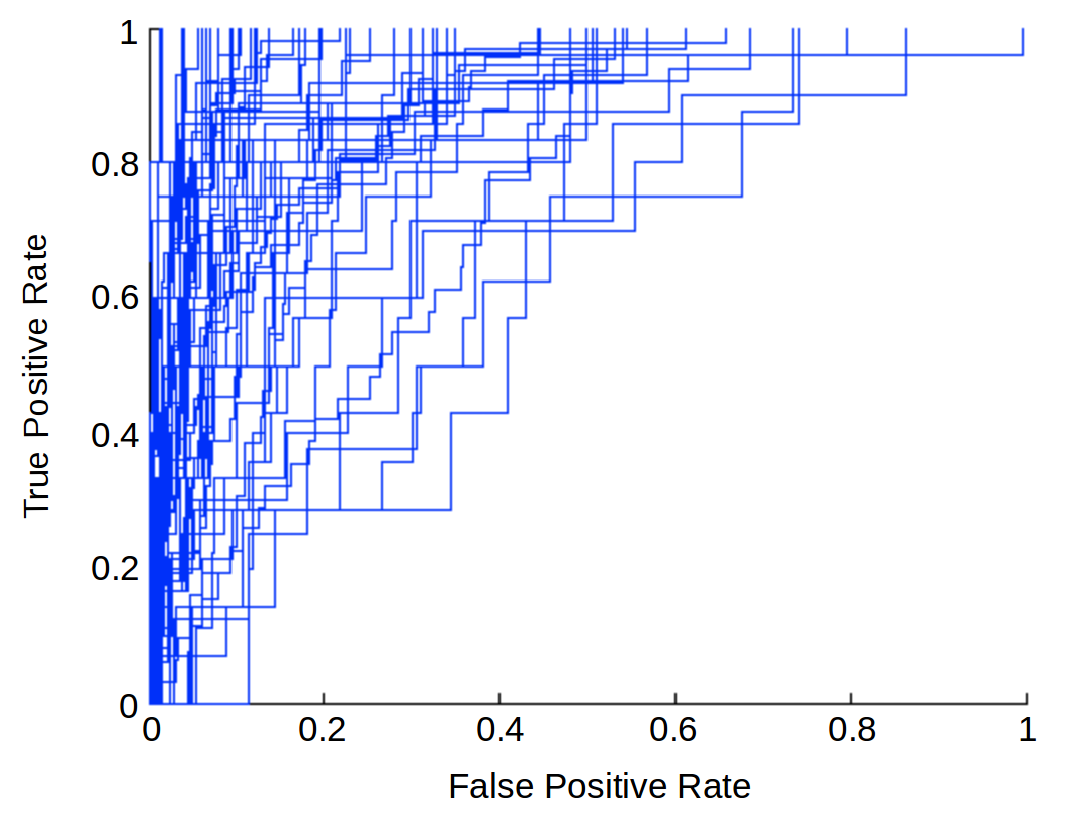}
\caption{The ROC curves corresponding to the 55 tasks. Plot was generated using  LIBSVM  tools \cite{Chang:2011}
}
\label{ROC}
\end{figure}

A protein superfamily is the largest collection of proteins for which a common ancestor can be traced. Within a superfamily, similarity between amino acid sequences may not be easily observed. Therefore, a superfamily can be further divided into several families within which, similarity among amino acid sequences usually can be identified. Members in a protein superfamily share similar structure with the common ancestor though they may not have similar sequences. In this case, based on structure information, we test our method in classification of protein superfamilies. The samples in this test case were taken from Protein Classification Benchmark Collection \cite{Sonego:2007}. The problem used in our test has the accession number of PCB00019. The goal of this data set is to classify protein domain sequences and structures into protein superfamilies, based on protein families. It contains 1357 samples and 55 classification tasks. Detailed description and classification results using different scoring method and various classification methods can be found in \href{http://hydra.icgeb.trieste.it/benchmark/index.php?page=.}{Protein Classification Benchmark Collection website}. In this test, we utilize  only the structure information of $\alpha$-carbon in protein backbones. For each task, we perform 5 fold cross validation on the training set to search for reasonable parameters. In most tasks, the numbers of positive and negative sets are unbalanced. To prevent unbalanced training results, different values for penalty parameters are used. Specifically, the   ratio between positive penalty parameter and negative penalty parameter is set to equal the ratio between number of negative instances and number of positive instances. The average accuracy for the positive testing set and the negative testing set are 82.29\% and 80.94\%, respectively. The average AUC value for the 55 tasks is 0.8954. The standard derivation of AUC for the 55 tasks is 0.09. Figure \ref{ROC} shows plots of the ROC curve for the 55 tasks.

\section{Discussion and Conclusion}\label{sec:Conclusion}

Persistent homology is a unique tool in computational topology and computational geometry. It explores the topology space by studying the evolution of simplicial complex over a filtration process of a given data set.  A nested sequence  of subsets are obtained by continuously increasing the filtration parameter. During the filtration process, birth and death of topological invariants  are recorded. The lifespan of a topological invariant shows how significant it is geometrically. Persistent homology is capable of discovering the  underlying topological feature of the space of interest and recognizing topologically small events. In other words, it gives not only information of global and significant topological features, but also perspective of local features of the underlying space. Persistent homology has been applied to computer graphics, geometric modeling, data analysis, and many other fields. A protein structure can be represented as point cloud in three dimension for atoms or graph with edges corresponding to different types of chemical bonds. This geometric nature of protein structures allows the application of persistent homology. In this work, we introduce the use of protein topological features captured by persistent homology for the protein classification. Our goal is to illustrate that molecular topological fingerprints (MTFs) can describe the structure of a protein from different perspectives and in different scales. This property of MTFs makes it possible to be used in protein classification from the topological point of view. We examine the performance of MTFs in several protein classification tasks with different emphasizes. We show that MTFs are a potential option for protein classification.

To introduce the topological features we used in classification, we briefly reviewed the definition of simplex and different types of simplicial complex. Basic concepts of filtration and persistence was recalled. We use $\alpha$-carbon atoms in M2 proton channel of influenza A to illustrate the filtration of a simplicial complex. We also showed the barcode plots for M2 channel in an  all-atom model and an $\alpha$-carbon model. Comparing these approach, it can be seen that all-atom model contains too many details which flood away useful information such as Betti 1 barcode representing alpha helices. Essentially, at the all-atom scale, different proteins have some common features due to the structures of  amino acids. Using a coarse-grained model with $\alpha$-carbon atoms reveals more information of the entire structure of the protein and dramatically reduces spatial complex and computational time. Therefore, we adapt coarse-grained model throughout this work. In some physical descriptions of proteins, an all-atom model may be preferred.

In persistent homology,  a convention is to cherish  long-persistent topological features which are presented as long-lived  bars in a barcode representation. Whereas, short-lived  barcodes are typically discharged as noise. In our case, the MTFs of proteins carry both global features and local traits. For protein analysis, both  global features and local traits are equally important.   In other words, it takes both long-lived topological features and short-lived topological traits to effectively characterize different proteins. A fundamental reason is that biomolecular structure, function, dynamics and transport are governed by the interactions of wide range of scales,  which
lead to multiple  characteristic length scales ranging from covalent bond, residue, secondary structure and domain dimensions to protein sizes. Based on our understanding of protein  characteristic length scales \cite{KLXia:2015c}, we are able to identify the responding protein topological fingerprints and determine their relevance and importance in protein classification.

To use of MTFs for the analysis of large scale biomoleculoar data, we have developed persistent homology based machine learning method. Essentially, we construct feature vectors by using MTFs. We utilize the support vector machine (SVM) algorithm, which is known for its robustness and high accuracy,  in our study.  The resulting MTF-SVM classifier is validated by four test cases. First, we explore the performance of the present MTF-SVM classifier for distinguishing drug bound M2 channels of influenza A virus from those of nature M2 channels. It is found that the proposed method does an excellent job in analyzing viral drug inhibition. A 96\% prediction accuracy is recorded.
In our second test, we consider  the  discrimination of hemoglobin molecules in their relaxed and taut forms. Again, the present approach works very well (80\% accuracy) for this problem.
 We further employ our MTF-SVM classifier for the identification of all alpha, all beta, and alpha-beta protein domains.  A total of 900 proteins is used in our study. Due to the relatively large sample size, a  5-fold cross-validation was carried out to optimize training parameters and validate the present method.  In this study,  the detailed local topological features facilitate the classification of proteins with different secondary structures. An average of  85\% accuracy is found over three protein classes. Finally, we utilize the present method for the classification of protein superfamilies.  We adapt a standard test, accession number   PCB00019,  from  Protein Classification Benchmark Collection \cite{Sonego:2007}.  It involves 1357 samples and 55 classification tasks.   A combination of both local and global topological features enables us to separate protein superfamilies. Based on 5-fold cross validation, an average classification accuracy of 82\% is found.

The objective of the present work is to examine the utility, accuracy and efficiency of computational topology for protein classification. As such, only topological information
is employed. The extensive test study establishes topology as an independent and valuable option for  large scale protein classification. Obviously, the present method can be improved in  a variety of ways. Specifically, one can combine topological features with other more established features, namely, sequence features and physical features. for protein analysis and classification. Indeed, MTFs computed from persistent homology differ sharply from sequence and physical based features. Therefore, a combination of topological features, sequence features, and physical features  must be able to take advantages of these three classes of methods. This aspect is beyond the scope of the present work and will be explored in our future research.

In our earlier work, we have introduced computational topological for mathematical modeling and prediction, such as molecular stability prediction \cite{KLXia:2014c}, protein folding analysis \cite{KLXia:2015e}, and protein bond length prediction \cite{KLXia:2015c}. The present work indicates that the combination of machine learning and computational topology will create a new powerful approach topology based mathematical modeling and prediction.

\vspace{1cm}
\section*{Acknowledgments}

This work was supported in part by NSF grants   IIS-1302285,  and DMS-1160352, and NIH Grant R01GM-090208. GWW thanks the Mathematical Biosciences Institute (MBI) for generous  financial support.

%\vspace{1cm}
%\bibliographystyle{abbrv}
%%\bibliography{refs}
%%\bibliography{../refs}
%
%\begin{thebibliography}{10}
%
%\bibitem{Allen:1987}
%M.~P. Allen and D.~J. Tildesley.
%\newblock {\em Computer Simulation of Liquids}.
%\newblock Oxford: Clarendon Press, 1987.
%
%
%\end{thebibliography}
\newpage

%\bibliography{refs}
%\bibliographystyle{unsrt}
%\bibliographystyle{abbrv}
%

\newpage

\appendix{Appendix: Instances used in Section \ref{sec:Case3} }

In this appendix we list protein   SCOPe IDs used in Section \ref{sec:Case3}.

\begin{table}[ht]
\footnotesize
\centering
\caption{All alpha proteins}
\begin{tabular}{|c|c|c|c|c|c|c|c|c|c|}
\hline
d1ux8a\_ & d3p4pb2 & d1grja1 & d2zjrv1 & d2qwob\_ & d1fxkc\_ & d1cxzb\_ & d1seta1 & d1k4ta1 & d1qoja\_ \\
d2jdih1 & d1uera1 & d1pv0a\_ & d1rfya\_ & d1tjla1 & d1x4ta1 & d2a26a1 & d2f6ma\_ & d1z0pa1 & d1z0jb1 \\
d2g0ua1 & d2hepa1 & d1gu2a\_ & d9anta\_ & d1sfea1 & d1c20a\_ & d1biaa1 & d1opca\_ & d1hc8a\_ & d2qanr1 \\
d1whua\_ & d1k6ya1 & d1twfj\_ & d1jhga\_ & d1ku2a1 & d1vz0a1 & d1rq6a\_ & d1cuka1 & d1veja1 & d1r5la1 \\
d1enwa\_ & d1eija\_ & d1sr9a1 & d1t95a1 & d1ufza\_ & d3ugja2 & d1jjcb1 & d1quua1 & d2e2aa\_ & d1jnra1 \\
d1g73a\_ & d1qsda\_ & d2qant1 & d3ldqb\_ & d1j2jb\_ & d1rrza\_ & d1z8ua\_ & d1vcta1 & d1vp7a\_ & d1wfda\_ \\
d1xdpa1 & d2crba1 & d2goma1 & d1lp1a\_ & d1gvna\_ & d3bvua1 & d1u00a1 & d1oksa\_ & d1nu9c1 & d1r8ia\_ \\
d2qkwa1 & d2ahma1 & d2gf4a1 & d2oo2a1 & d1ebdc\_ & d2erla\_ & d1x9ba\_ & d1hb6a\_ & d1gg3a1 & d4i9oa\_ \\
d2fcwa1 & d1ujsa\_ & d1tbaa\_ & d1aila\_ & d2hp8a\_ & d2enda\_ & d3lyna\_ & d2wkxa1 & d2gzka1 & d1kx5c\_ \\
d1fpoa2 & d1eexg\_ & d1mtyg\_ & d1om2a\_ & d1jw2a\_ & d2gboa1 & d2gsva1 & d1nfoa\_ & d2asra\_ & d256ba\_ \\
d1i4ya\_ & d1cgme\_ & d3fapb\_ & d1ya7o1 & d1h6ga1 & d2a0ba\_ & d1he1a\_ & d1f1ma\_ & d1wgwa\_ & d1qvxa\_ \\
d3tk0a\_ & d1knya1 & d1l3pa\_ & d1nzea\_ & d1orja\_ & d1v74b\_ & d1gm5a1 & d1t6ua\_ & d1szia\_ & d1ug7a\_ \\
d1xzpa1 & d2huja1 & d2ap3a1 & d2g3ka1 & d2p61a1 & d4lqha\_ & d1niga\_ & d1wa8a1 & d2g38a1 & d2gtsa1 \\
d2nr5a1 & d1cnt1\_ & d1f7ua1 & d2af8a\_ & d1unka\_ & d2eiaa1 & d4oufa\_ & d1u8va1 & d1gkza1 & d1rj1a\_ \\
d1r3ba\_ & d1tdpa\_ & d2etda1 & d1v9va1 & d2fefa1 & d2fug11 & d3dbya1 & d2qzga1 & d2hi7b1 & d2rlda1 \\
d2hgka1 & d1nkda\_ & d1joya\_ & d1pd3a\_ & d1ufia\_ & d1skva\_ & d1zkea1 & d2bzba1 & d2az0a1 & d3im3a\_ \\
d1nh2b\_ & d1ecia\_ & d1b0nb\_ & d1g2ya\_ & d1pzqa\_ & d1q2ha\_ & d1ic8a2 & d1hq1a\_ & d1k1va\_ & d1nlwa\_ \\
d1pzra\_ & d1qx2a\_ & d2jpoa\_ & d2ciwa1 & d1iioa\_ & d1sh5a1 & d1lnsa1 & d1wixa\_ & d1a26a1 & d1v32a\_ \\
d1baza\_ & d1k0ma1 & d3bula1 & d1khda1 & d2fzta1 & d1uura1 & d1fioa\_ & d1s2xa\_ & d3nyla\_ & d2fupa1 \\
d1t98a2 & d3buxb2 & d3bi1a1 & d1wjta\_ & d2b4jc1 & d2okua1 & d1f6va\_ & d2a73b3 & d1v54h\_ & d1n89a\_ \\
d1aiea\_ & d1adua1 & d1p71a\_ & d1rm6c1 & d1dj8a\_ & d1af7a1 & d2rmra\_ & d1sv0a\_ & d1cuka2 & d1z3eb1 \\
d3ldaa1 & d1ci4a\_ & d4klia1 & d1bgxt1 & d1d8ba\_ & d1f44a1 & d1zyma1 & d1ryka\_ & d4klia2 & d1m6ya1 \\
d1q46a1 & d2fj6a1 & d3ci0k1 & d2gola\_ & d1qgta\_ & d1aepa\_ & d1l9la\_ & d1o82a\_ & d1n00a\_ & d1tada1 \\
d1ej5a\_ & d1skyb1 & d1fkma1 & d1q0qa1 & d2o3la1 & d1abva\_ & d2oeba1 & d1g7da\_ & d2qtva1 & d1dvka\_ \\
d1u7ka\_ & d3o0gd\_ & d1husa\_ & d2qq9a2 & d3ezqb\_ & d1hw1a2 & d1ey1a\_ & d1a5ta1 & d1b79a\_ & d3ju5a1 \\
d1tx9a1 & d1llaa1 & d1llaa2 & d1by1a\_ & d1boua\_ & d1hbna1 & d1bgfa\_ & d1dk8a\_ & d1a9xa1 & d1apxa\_ \\
d1vq8p1 & d1aa7a\_ & d2abka\_ & d1j09a1 & d1rlra1 & d1dnpa1 & d2pgda1 & d1zkra1 & d1gaia\_ & d1dl2a\_ \\
d1qaza\_ & d1n1ba1 & d1r76a\_ & d2g0da\_ & d1a59a\_ & d1io7a\_ & d1rqta\_ & d1iiea\_ & d1aora1 & d1d2ta\_ \\
d1wb9a1 & d1f5na1 & d1bvp11 & d1xa6a1 & d1nvus\_ & d1jdha\_ & d1re0b\_ & d1lsha1 & d1qsaa1 & d2h6fa1 \\
d2o8pa1 & d1ihga1 & d1inza\_ & d3ag3e\_ & d2grrb\_ & d1k8kg\_ & d4fhrb\_ & d1l5ja1 & d1j1ja\_ & d1ldja2 \\
\hline
\end{tabular}
\label{tab:allalpha}
\end{table}

\begin{table}[ht]
\footnotesize
\centering
\caption{All beta proteins}
\begin{tabular}{|c|c|c|c|c|c|c|c|c|c|}
\hline
d2giya1 & d1x4za1 & d1b4ra\_ & d1yq2a1 & d1ex0a2 & d1l3wa1 & d1acxa\_ & d1ej8a\_ & d1pl3a\_ & d1kyfa1 \\
d1p5va1 & d1kbpa1 & d1vzia1 & d1f00i1 & d1tyeb1 & d1ifra\_ & d1l6pa\_ & d2cxka1 & d1lmia\_ & d1o75a1 \\
d1osya\_ & d1roca\_ & d1xq4a\_ & d1xaka\_ & d1xo8a\_ & d2nqda\_ & d2dpka1 & d2itea1 & d1edya\_ & d4qmea3 \\
d3abza4 & d1f0la1 & d2xbda\_ & d1amxa\_ & d3d06a\_ & d1e2wa1 & d1h6ea\_ & d2qtva2 & d2dexx1 & d2jqaa1 \\
d1vema1 & d1h8la1 & d1lm8v\_ & d1tfpa\_ & d1d2oa1 & d1dmha\_ & d1xpna\_ & d1c3ga1 & d1ok0a\_ & d2ov0a\_ \\
d1kzqa1 & d1rlwa\_ & d1p5va2 & d1n10a1 & d1dcea2 & d2huha1 & d3ivva\_ & d2hnua\_ & d2bb2a1 & d1loxa2 \\
d1kful2 & d4elda\_ & d1slua\_ & d1wpxb1 & d1gofa2 & d1bvp12 & d1aola\_ & d2j1kc1 & d1c3ha\_ & d1sfpa\_ \\
d2f4la1 & d2a73b6 & d1hn0a3 & d1pcva\_ & d1khua\_ & d1cq3a\_ & d1p35a\_ & d1viwb\_ & d1w6ga1 & d1yq2a4 \\
d1jmma\_ & d1h2ca\_ & d1s2ea\_ & d1g8kb\_ & d1biaa2 & d1ycsb2 & d1kk8a1 & d1viea\_ & d1vq8q1 & d3vuba\_ \\
d1ex4a1 & d1hyoa1 & d2qqra1 & d2coza1 & d1m9sa2 & d3ptaa3 & d1azpa\_ & d1r4ka\_ & d1sf9a\_ & d1y71a1 \\
d1vbva1 & d2eyqa1 & d3dkma\_ & d2heqa1 & d3u1ua\_ & d1pcqo\_ & d1l1ca\_ & d1t2ma1 & d2g3pa1 & d1i4k1\_ \\
d1ib8a1 & d2d6fa1 & d2qi2a1 & d2f5tx1 & d1whia\_ & d4dfaa\_ & d3chbd\_ & d3v96a\_ & d1eova1 & d1e9ga\_ \\
d1guta\_ & d2ch4w1 & d2p5zx1 & d1sr3a\_ & d1nnxa\_ & d2f4ia1 & d2exda1 & d2k5wa1 & d2ot2a1 & d4n8x1\_ \\
d3rloa1 & d2j8ch2 & d1bara\_ & d2zqoa1 & d1jlya1 & d3bx1c\_ & d3llpa1 & d1t9fa\_ & d3brda3 & d1wd3a2 \\
d1fuia1 & d1n0ua1 & d1ja1a1 & d1n08a\_ & d1f60a2 & d1yx2a2 & d2fg9a1 & d1ywua1 & d2jn9a1 & d2bw0a1 \\
d1arba\_ & d1bcoa1 & d1skyb2 & d1bd0a1 & d1yloa1 & d1fmba\_ & d1ffya2 & d1wc2a1 & d1ppya\_ & d1dfup\_ \\
d1h9db\_ & d1w1ha\_ & d1ywya1 & d1nh2c\_ & d1o6ea\_ & d1a3wa1 & d1ik9a1 & d1i4ua\_ & d1nqna\_ & d1smpi\_ \\
d1ei5a1 & d1pbya5 & d2djfa\_ & d1y0ga\_ & d2gtln1 & d2f09a1 & d2c3ba1 & d1a1xa\_ & d2rl8a\_ & d1f3ub\_ \\
d1su3a2 & d1tl2a\_ & d1gyha\_ & d1s1da\_ & d1v3ea\_ & d1crua\_ & d1h6la\_ & d2hqsa1 & d1ijqa1 & d3o4pa\_ \\
d1k32a2 & d1ofza\_ & d1q7fa\_ & d1suua\_ & d1zgka1 & d1gofa3 & d2bbkh\_ & d1fwxa2 & d1pgua1 & d1a12a\_ \\
d3gc3b1 & d2xdwa1 & d3niga\_ & d1k32a3 & d1jofa\_ & d1ri6a\_ & d1shyb1 & d1sqja1 & d1xksa\_ & d1flga\_ \\
d1qksa2 & d1xfda1 & d1m7xa2 & d2ho2a1 & d1aiwa\_ & d1rk8c\_ & d1dkga1 & d1v9ea\_ & d3enia\_ & d1ospo\_ \\
d2f69a1 & d1vmoa\_ & d1i5pa2 & d1vboa\_ & d1dlpa1 & d1pxza\_ & d1ezga\_ & d1hf2a1 & d1ea0a1 & d1k4za\_ \\
d1vh4a\_ & d1k7ia1 & d2bm4a1 & d2qiaa\_ & d1l0sa\_ & d1p9ha\_ & d2cu2a1 & d1ep0a\_ & d1w9ya1 & d1ft9a2 \\
d2arca\_ & d1gtfa\_ & d1ig0a1 & d3ar4a2 & d1v1ha1 & d1dcza\_ & d1w96a1 & d1gpra\_ & d2zjrt1 & d2zdra1 \\
d1c5ea\_ & d4ubpb\_ & d1pkha\_ & d1tula\_ & d2ftsa1 & d1ml9a\_ & d1at0a\_ & d1f39a\_ & d2fu5a1 & d2rqaa\_ \\
d3ezma\_ & d1lkta\_ & d1r6na\_ & d4ubpc1 & d2jdih2 & d1f35a\_ & d2ag4a\_ & d3zdha\_ & d1iaza\_ & d4qmea1 \\
d2oqza\_ & d1e44b\_ & d1fjra\_ & d2ftsa2 & d1jhna3 & d1xp4a1 & d1wrua1 & d1eara1 & d1h6wa1 & d1h09a1 \\
d1rh1a1 & d1p6va\_ & d1js8a2 & d3gpua1 & d1m1ha1 & d4ce8a\_ & d1mkfa\_ & d1lnza1 & d1o70a1 & d1ko6.1 \\
d1o75a3 & d1pgsa1 & d1hx6a1 & d1nlqa\_ & d1qqp.1 & d1m06f\_ & d1dzla\_ & d1stma\_ & d1iq8a3 & d1nc7a\_ \\
\hline
\end{tabular}
\label{tab:allbeta}
\end{table}

\begin{table}[ht]
\footnotesize
\centering
\caption{Alpha+beta proteins}
\begin{tabular}{|c|c|c|c|c|c|c|c|c|c|}
\hline
d2c4bb1 & d2baaa\_ & d1qdqa\_ & d2jb0b\_ & d1km8a\_ & d1xu0a\_ & d1y7ma2 & d4ubpa\_ & d1el0a\_ & d2cs7a1 \\
d1bb8a\_ & d1qmea1 & d2zjrq1 & d1fita\_ & d2r7ja1 & d1n0ua3 & d1v5oa\_ & d1ibxa\_ & d1v8ca1 & d1czpa\_ \\
d2saka\_ & d2g9hd2 & d1hz6a\_ & d1tifa\_ & d1htqa1 & d1tkea1 & d1mjda\_ & d3n20c\_ & d2fug13 & d2gria1 \\
d3coxa2 & d1mola\_ & d1w6ga2 & d1eeja2 & d1idpa\_ & d1udii\_ & d1iq8a4 & d1nnva\_ & d1pcfa\_ & d4f7ea1 \\
d2f4za1 & d4ijza1 & d1xqma\_ & d1c8za\_ & d2pila\_ & d1p32a\_ & d1ix5a\_ & d3eipa\_ & d1w9pa2 & d3bn0a1 \\
d2qans1 & d1vq8x1 & d1b33n\_ & d1cr5a2 & d1f9za\_ & d1qyna\_ & d1bm8a\_ & d1dk0a\_ & d1eyqa\_ & d2v8qe1 \\
d1lo7a\_ & d3bria\_ & d1to2i\_ & d1rm6a1 & d2i1oa1 & d1brwa3 & d1vq8h1 & d1fm0e\_ & d1r29a\_ & d1xb2b2 \\
d1wina\_ & d1uera2 & d2zjq51 & d1wiba\_ & d1xp8a2 & d4ghla\_ & d1e8ob\_ & d1whqa\_ & d1pdaa2 & d1dq3a2 \\
d1j26a\_ & d2fgga1 & d2z0sa2 & d3proc1 & d1gpma3 & d3ieua2 & d1ib8a2 & d1mkya3 & d1v9ja\_ & d1kkga\_ \\
d1veha\_ & d2qfia1 & d2bh1x1 & d2qanc2 & d1kkoa2 & d2zjrp1 & d1kp8a3 & d1ghha\_ & d1blua\_ & d4fyxb1 \\
d1jqga2 & d1mlia\_ & d2cz4a1 & d3b6ba\_ & d2cq2a1 & d1b3ta\_ & d1bxna2 & d2acya\_ & d1n0ua4 & d1gh8a\_ \\
d1jjcb4 & d2j5aa1 & d2qanj1 & d2q66a3 & d1cc8a\_ & d1sc6a3 & d1z2la2 & d1dqaa1 & d1ekra\_ & d1f3va\_ \\
d1mlaa2 & d1ffgb\_ & d1kp6a\_ & d1h72c2 & d1regx\_ & d1fvga\_ & d1wc3a\_ & d3hd2a\_ & d1hbnc\_ & d1diqa1 \\
d1m5ha1 & d1qd1a1 & d2akja1 & d1eara2 & d1gpja3 & d1o8ba2 & d1ivza\_ & d1m1ha2 & d2vv5a2 & d1oy8a1 \\
d1pbua\_ & d1nxia\_ & d1yqha1 & d1in0a1 & d1j27a\_ & d1q8ka2 & d1utaa\_ & d1wj9a1 & d1rwua\_ & d2p8ia\_ \\
d2a1ba1 & d2vjva1 & d4tnoa\_ & d2av5a1 & d3b8pa1 & d3bpda1 & d1zava1 & d1vq8w1 & d1r8ea2 & d1jh6a\_ \\
d1f32a\_ & d1d8ia\_ & d2if1a\_ & d2p92a1 & d1lbua2 & d1c05a\_ & d1tkea2 & d1f7ua3 & d1ge9a\_ & d1uv7a\_ \\
d1tiga\_ & d1qmha2 & d1pava\_ & d1rq8a\_ & d1nj8a2 & d1nfja\_ & d1ug8a\_ & d2d9ia1 & d1kpta\_ & d1ev0a\_ \\
d1dwka2 & d1bxni\_ & d1ru0a\_ & d1b4ba\_ & d1bdfa1 & d2d6fc2 & d2phcb2 & d1r0va3 & d1wb9a4 & d1syxb\_ \\
d2ckca1 & d1iq4a\_ & d1dzfa2 & d1nq3a\_ & d2r6r12 & d1t0kb\_ & d1clia1 & d3erna\_ & d1q8ra\_ & d1r1ma\_ \\
d2ohwa1 & d3byqa1 & d1bjpa\_ & d1nvmb2 & d2iafa1 & d2glza1 & d2dm9a1 & d1oaca4 & d4ec2a\_ & d1sgoa\_ \\
d1v5ra1 & d1ylxa1 & d1ewfa1 & d1usub\_ & d2sici\_ & d1frsa\_ & d1rf8a\_ & d1m6ia3 & d1rm6b1 & d1mnma\_ \\
d4nbpa\_ & d1vfra\_ & d1dt9a3 & d1c7ka\_ & d1qbaa4 & d2abla2 & d1mo1a\_ & d1j5ya2 & d3bp3a\_ & d2qojz1 \\
d1a8ra\_ & d1vrma1 & d1qb3a\_ & d1jtgb\_ & d3elga1 & d1diva1 & d2hbaa1 & d2gpfa1 & d2ba0d2 & d1efnb\_ \\
d1cbya\_ & d1yfsa2 & d1kyfa2 & d1wfra\_ & d1jhsa\_ & d1m4ia\_ & d1d0na1 & d2qtva4 & d2gnxa2 & d1acfa\_ \\
d1mc0a1 & d1nwza\_ & d1ifqa\_ & d1l3la2 & d1ojga\_ & d1z09a\_ & d2h28a1 & d2a2la1 & d2grga1 & d1rc9a1 \\
d1a6ja\_ & d4n1ta\_ & d1hp1a1 & d1jcua\_ & d1wdva\_ & d1qzfa2 & d2y28a\_ & d1cyoa\_ & d1vcca\_ & d2iwxa\_ \\
d1ixma\_ & d1iooa\_ & d1c4ka3 & d1g62a\_ & d2v3za2 & d1htqa2 & d1ytba1 & d1kfia4 & d2q3qa\_ & d1f46a\_ \\
d1v8ca2 & d1ul7a\_ & d1j3ma\_ & d1t6aa\_ & d1xsza2 & d2fpna1 & d2pwwa1 & d1mxaa1 & d1ok7a1 & d1rm6a2 \\
d3mm5a2 & d1go4a\_ & d1byra\_ & d1hqia\_ & d1clia2 & d1seia\_ & d1rl6a1 & d1iowa2 & d1a0ia2 & d2cnqa\_ \\
\hline
\end{tabular}
\label{tab:aplb}
\end{table}

\end{document}